\documentclass[aps,prd,twocolumn,superscriptaddress]{revtex4}
\usepackage{epsfig,epsf}
\usepackage{amsmath}
\usepackage{amsthm}
\usepackage{amsfonts}
\usepackage{amssymb}
\usepackage{dsfont}
\usepackage{multirow}
\usepackage{appendix}
\usepackage{slashed}
\usepackage[active]{srcltx}
\usepackage{psfrag}

\begin{document}

\title{Resonance $X(3960)$ as a hidden charm-strange scalar tetraquark}
\date{\today}
\author{S.~S.~Agaev}
\affiliation{Institute for Physical Problems, Baku State University, Az--1148 Baku,
Azerbaijan}
\author{K.~Azizi}
\affiliation{Department of Physics, University of Tehran, North Karegar Avenue, Tehran
14395-547, Iran}
\affiliation{Department of Physics, Do\v{g}u\c{s} University, Dudullu-\"{U}mraniye, 34775
Istanbul, Turkey}
\author{H.~Sundu}
\affiliation{Department of Physics, Kocaeli University, 41380 Izmit, Turkey}

\begin{abstract}
We investigate features of the hidden charm-strange scalar tetraquark $c%
\overline{c}s\overline{s}$ by calculating its spectral parameters and width,
and we compare the obtained results with the mass and width of the resonance $%
X(3960)$ discovered recently in the LHCb experiment. We model the tetraquark
as a diquark-antidiquark state $X=[cs][\overline{c}\overline{s}]$ with
spin-parities $J^{\mathrm{PC}}=0^{++}$. The mass and current coupling of $X$
are calculated using the QCD two-point sum rules by taking into account
various vacuum condensates up to dimension $10$. The width of the tetraquark
$X$ is estimated via the decay channels $X \to D_{s}^{+}D_{s}^{-}$ and $X \to
\eta_{c} \eta^{(\prime)}$. The partial widths of these processes are
expressed in terms of couplings $G$, $g_1$ and $g_2$ which describe the strong
interactions of particles at the vertices $XD_{s}^{+}D_{s}^{-}$, $%
X\eta_{c}\eta^{\prime}$ and $X\eta_{c}\eta$, respectively. Numerical values
of $G$, $g_1$ and $g_2$ are evaluated by employing the three-point sum rule
method. Comparing the results $m=(3976 \pm 85)~\mathrm{MeV}$ and $\Gamma_{%
\mathrm{X}}=(42.2 \pm 12.0)~\mathrm{MeV}$ obtained for parameters of the
tetraquark $X$ and experimental data of the LHCb Collaboration, we conclude
that the resonance $X(3960)$ can be considered as a candidate to a scalar
diquark-antidiquark state.
\end{abstract}

\maketitle

%%%%%%%%%%%%%%%%%%%%%%%%%%%%%%%%%%%%%%%%%%%%%%%%%%%%%%%%%%%%%%%%

\section{Introduction}

\label{sec:Int}
%%%%%%%%%%%%%%%%%%%%%%%%%%%%%%%%%%%%%%%%%%%%%%%%%%%%%%%%%%%%%%%%%%%%%%%

Recently, the LHCb Collaboration reported the observation of a new threshold
peaking structure $X(3960)$ in the $D_{s}^{+}D_{s}^{-}$ invariant mass
distribution in the $B^{+}\rightarrow D_{s}^{+}D_{s}^{-}K^{+}$ decay \cite%
{LHCb:2022vsv}. Performed analysis demonstrated that it is a scalar
resonance $J^{\mathrm{PC}}=0^{++}$ with the mass and width
\begin{eqnarray}
m_{\mathrm{exp} } &=&3956\pm 5\pm 10~\mathrm{MeV},  \notag \\
\Gamma _{\mathrm{exp} } &=&43\pm 13\pm 8~\mathrm{MeV}.  \label{eq:Data}
\end{eqnarray}%
The collaboration also found an additional structure around $4140\ \mathrm{%
MeV}$ with spin-parities $0^{++}$. The resonance $X(3960)$ was interpreted
by LHCb as a four-quark state with the content $c\overline{c}s\overline{s}$,
whereas the structure $4140~\mathrm{MeV}$ may be either a new resonance or $%
J/\psi \phi \leftrightarrow D_{s}^{+}D_{s}^{-}$ coupled-channel effect.

Four-quark exotic mesons composed of quarks $c\overline{c}s\overline{s}$
with different quantum numbers are not something new for both experimental
and theoretical physicists. In fact, resonances with the quark content $c%
\overline{c}s\overline{s}$ were fixed by LHCb in the $J/\psi \phi $
invariant mass distribution in the process $B^{+}\rightarrow J/\psi \phi
K^{+}$ \cite{Aaij:2016iza}. The discovered states $X(4140)$ and $X(4274)$
are axial-vector particles with $J^{\mathrm{PC}}=1^{++}$, whereas the
spin-parities of $X(4500)$ and $X(4700)$ are $J^{\mathrm{PC}}=0^{++}$. It
should be noted that resonances $X(4140)$ and $X(4274)$ were previously seen
by the CDF Collaboration \cite{Aaltonen:2009tz} in the decays $B^{\pm
}\rightarrow J/\psi \phi K^{\pm }$, and confirmed later by CMS \cite%
{Chatrchyan:2013dma} and D0 experiments \cite{Abazov:2013xda}. The scalar
structures $X(4500)$ and $X(4700)$ were fixed by the LHCb Collaboration for
the first time.

In experiments numerous exotic vector mesons built of $c\overline{c}s%
\overline{s}$ quarks were observed as well. Thus, the state $Y(4660)$ was
found for the first time by the Belle Collaboration in the process $%
e^{+}e^{-}\rightarrow \gamma _{\mathrm{ISR}}\psi (2S)$ $\pi ^{+}\pi ^{-}$ as
one of two resonant structures in the $\psi (2S)\pi ^{+}\pi ^{-}$ invariant
mass distribution. Because $Y(4660)$ was produced in the $e^{+}e^{-}$
annihilation its quantum numbers are $J^{\mathrm{PC}}=1^{--}$. The structure
$X(4630)$ was discovered by LHCb in the $J/\psi \phi $ invariant mass
distribution of the decay $B^{+}\rightarrow J/\psi \phi K^{+}$ \cite%
{LHCb:2021uow}.

Theoretical studies of four-quark states $c\overline{c}s\overline{s}$ have
also rich history. The charmoniumlike exotic mesons with $s\overline{s}$
component were investigated by means of different methods in numerous
publications (see, as examples, Refs.\ \cite%
{Nieves:2012tt,Wang:2013exa,Lebed:2016yvr,Chen:2017dpy,Meng:2020cbk}).
Comprehensive analyses of some of hidden-charm diquark-antidiquark systems $%
[cs][\overline{c}\overline{s}]$ were carried out in our articles as well.
Thus, the axial-vector resonances $X(4140)$ and $X(4274)$ were investigated
in Ref.\ \cite{Agaev:2017foq}, in which, we treated them as
diquark-antidiquark states built of scalar and axial-vector components
belonging to triplet and sextet representations of $SU_{c}(3)$ color group,
respectively. We calculated not only their masses and current couplings (or
pole residues) but also evaluated full widths of these tetraquarks.
Predictions for parameters of \ the color-triplet diquark-antidiquark state
allowed us to interpret it as the resonance $X(4140)$. Contrary, the full
width of the tetraquark with color sextet ingredients is considerably wider
than that of the resonance $X(4274)$. Therefore, to explain the internal
organization of $X(4274)$ alternative models should be examined, though
existence of a new axial-vector resonance with the mass $m\approx 4274~%
\mathrm{MeV}$ and full width $\Gamma \approx 200~\mathrm{MeV}$ cannot be
excluded.

The vector resonance $Y(4660)$ was studied as a diquark-antidiquark vector
state $[cs][\overline{c}\overline{s}]$ with $J^{\mathrm{PC}}=1^{--}$ in our
work \cite{Sundu:2018toi}. Results obtained there for the mass and full
width of this structure made it possible to interpret the resonance $Y(4660)$
as the diquark-antidiquark exotic meson. The detailed analysis of $X(4630)$
was performed in Ref.\ \cite{Agaev:2022iha} by assuming that it is a vector
tetraquark $[cs][\overline{c}\overline{s}]$ with spin-parities $J^{\mathrm{PC%
}}=1^{-+}$. Here, a nice agreement was obtained between the LHCb data for
parameters of the resonance $X(4630)$ and theoretical predictions of the
diquark-antidiquark model. There are numerous articles devoted to
experimental studies and theoretical analysis of hidden charm-strange
four-quark mesons in the literature: Relatively full list of such
publications can be found in Refs.\ \cite%
{LHCb:2022vsv,Agaev:2017foq,Sundu:2018toi,Agaev:2022iha}.

First announcement made in Ref.\ \cite{X3960} about discovery of the
resonance $X(3960)$ triggered extreme interest to this state. In papers \cite%
{Bayar:2022dqa,Ji:2022uie,Xin:2022bzt,Xie:2022lyw,Chen:2022dad,Guo:2022ggl,Guo:2022crh}
appeared afterwards, authors addressed different aspects of its internal
organization, production mechanisms and rates, placed $X(3960)$ into various
four-quark multiplets. The coupled-channel explanation of $X(3960)$ was
suggested in Ref.\ \cite{Bayar:2022dqa}, where it emerges as an enhancement
in the $D_{s}^{+}D_{s}^{-}$ mass distribution via interaction of the $%
D^{+}D^{-}$ and $D_{s}^{+}D_{s}^{-}$ coupled-channels. In Ref.\ \cite%
{Xin:2022bzt} the authors assigned $X(3960)$ the hadronic molecule $%
D_{s}^{+}D_{s}^{-}$, and performed studies in the context of the sum rule
method. The resonance $X(3960)$ was explained also as near the $%
D_{s}^{+}D_{s}^{-}$ threshold enhancement due to the contribution of the
conventional $P$-wave charmonium $\chi _{c0}(2P)$ \cite{Guo:2022ggl}.

In the present article, we explore the tetraquark $X=[cs][\overline{c}%
\overline{s}]$ with spin-parities $J^{\mathrm{PC}}=0^{++}$ and compute its
parameters. The mass and current coupling of $X$ are evaluated using the QCD
two-point sum rule method. Its full width is estimated using the decay
channels $X\rightarrow D_{s}^{+}D_{s}^{-}$ and $X\rightarrow \eta _{c}\eta
^{(\prime )}$. Partial widths of these processes are expressed through
strong couplings $G$, $g_{1}$ and $g_{2}$ of particles at the vertices $%
XD_{s}^{+}D_{s}^{-}$, $X\eta _{c}\eta ^{\prime }$, and $X\eta _{c}\eta $,
respectively. To calculate $G$, $g_{1}$ and $g_{2}$, we employ technical
tools of the three-point sum rule approach. Results found for parameters of
the state $X$ are confronted with the LHCb data to verify the
diquark-antidiquark model for $X(3960)$.

This paper is organized in the following way: In Sec.\ \ref{sec:Mass}, we
compute the mass and current coupling of the tetraquark $X$ by means of the
QCD two-point sum rule method. The decay $X\rightarrow D_{s}^{+}D_{s}^{-}$
is studied in Sec.\ \ref{sec:Decay1}, where we calculate the coupling $G$
and partial width of this process. The strong couplings $g_{1}$ and $g_{2}$
and partial widths of the decays $X\rightarrow \eta _{c}\eta ^{\prime }$ and
$X\rightarrow \eta _{c}\eta $, as well as the full width of $X$ are found in
Sec.\ \ref{sec:Decays2}. The section \ref{sec:Conclusion} is reserved for
our concluding notes.

%%%%%%%%%%%%%%%%%%%%%%%%%%%%%%%%%%%%%%%%%%%%%%%%%%%%%%%%%%%%%%%%%%%%%%%

\section{Mass and current coupling of the tetraquark $X$}

\label{sec:Mass}
%%%%%%%%%%%%%%%%%%%%%%%%%%%%%%%%%%%%%%%%%%%%%%%%%%%%%%%%%%%%%%%%%%%%%
In this section, we consider the scalar diquark-antidiquark state $X=[cs][%
\overline{c}\overline{s}]$ and extract its spectroscopic parameters from the
two-point sum rule analysis \cite{Shifman:1978bx,Shifman:1978by}. It is
known that the sum rule method operates with correlation functions and
interpolating currents of particles under investigations. There are
different ways to construct a scalar tetraquark and corresponding current
using a diquark and an antidiquark with different spin-parities \cite%
{Chen:2010ze}. Thus, one may construct such state using the pseudoscalar $%
c^{T}Cs$ or vector $c^{T}C\gamma _{\mu }\gamma _{5}s$ diquarks and
corresponding antidiquarks, where $C$ is the charge-conjugation operator.
But, we assume that $X$ is built of a scalar diquark $c^{T}C\gamma _{5}s$
and antidiquark $\overline{c}\gamma _{5}C\overline{s}^{T}$: The reason is
that the scalar diquark (antidiquark) configuration is the most attractive
and stable two-quark system \cite{Jaffe:2004ph}.

The structures $\epsilon c^{T}C\gamma _{5}s$ and $\widetilde{\epsilon }%
\overline{c}\gamma _{5}C\overline{s}^{T}$ are the color antitriplet and
triplet states of the color $SU_{c}(3)$ group, respectively. Then the
interpolating current for the tetraquark $X$ has the form

\begin{equation}
J(x)=\epsilon \widetilde{\epsilon }\left[ c_{b}^{T}(x)C\gamma _{5}s_{c}(x)%
\right] \left[ \overline{c}_{m}(x)\gamma _{5}C\overline{s}_{n}^{T}(x)\right]
,  \label{eq:C1}
\end{equation}%
where $\epsilon \widetilde{\epsilon }=\epsilon ^{abc}\epsilon ^{amn}$ and $%
a,b,c,m$ and $n$ are color indices. This current belongs to $[\overline{%
\mathbf{3}}_{c}]_{cs}\otimes \lbrack \mathbf{3}_{c}]_{\overline{c}\overline{s%
}}$ representation of the color group and corresponds to the scalar state
with quantum numbers $J^{\mathrm{PC}}=0^{++}$. The current $J(x)$ describes
a ground-state scalar particle with lowest mass and required spin-parities.

The mass $m$ and coupling $f$ of the tetraquark $X$ can be determined from
analysis of the correlation function $\Pi (p)$
\begin{equation}
\Pi (p)=i\int d^{4}xe^{ipx}\langle 0|\mathcal{T}\{J(x)J^{\dag
}(0)\}|0\rangle.  \label{eq:CF1}
\end{equation}%
To derive the required sum rules, one has to express $\Pi (p)$ using the
spectroscopic parameters of the tetraquark $X$. For these purposes, we
insert into the correlation function $\Pi (p)$ a complete set of states with
quantum numbers $0^{++}$ and perform integration over $x$ in Eq.\ (\ref%
{eq:CF1}). As a result, we get
\begin{equation}
\Pi ^{\mathrm{Phys}}(p)=\frac{\langle 0|J|X(p\rangle \langle X(p)|J^{\dagger
}|0\rangle }{m^{2}-p^{2}}+\cdots.  \label{eq:PhysSide}
\end{equation}%
The obtained expression forms a hadronic representation of $\Pi (p)$ and is the
phenomenological (physical) side of sum rule. Here, the  contribution coming
from the ground-state particle $X$ is written down explicitly, whereas
contributions of higher resonances and continuum states are denoted by
the ellipses.

The function $\Pi ^{\mathrm{Phys}}(p)$ can be further simplified by
employing the matrix element
\begin{equation}
\langle 0|J|X(p)\rangle =fm.  \label{eq:MElem1}
\end{equation}%
It is easy to find that, in terms of the parameters $m$ and $f$, the
function $\Pi ^{\mathrm{Phys}}(p)$ takes the following form
\begin{equation}
\Pi ^{\mathrm{Phys}}(p)=\frac{m^{2}f^{2}}{m^{2}-p^{2}}+\cdots.
\label{eq:PhysSide1}
\end{equation}%
The $\Pi ^{\mathrm{Phys}}(p)$ has simple Lorentz structure proportional to $%
\mathrm{I}$, and relevant invariant amplitude $\Pi ^{\mathrm{Phys}}(p^{2})$
is given by r.h.s. of Eq.\ (\ref{eq:PhysSide1}).

To determine the QCD side of the sum rules $\Pi ^{\mathrm{OPE}}(p)$, we use
the interpolating current $J(x)$ in Eq.\ (\ref{eq:CF1}), and contract the  heavy
and light quark fields. After simple manipulations, we obtain
\begin{eqnarray}
&&\Pi ^{\mathrm{OPE}}(p)=i\int d^{4}xe^{ipx}\epsilon \widetilde{\epsilon }%
\epsilon ^{\prime }\widetilde{\epsilon }^{\prime }\mathrm{Tr}\left[ \gamma
_{5}\widetilde{S}_{c}^{bb^{\prime }}(x)\right.  \notag \\
&&\left. \times \gamma _{5}S_{s}^{cc^{\prime }}(x)\right] \mathrm{Tr}\left[
\gamma _{5}\widetilde{S}_{s}^{n^{\prime }n}(-x)\gamma _{5}S_{c}^{m^{\prime
}m}(-x)\right] ,  \label{eq:QCDside}
\end{eqnarray}%
where $S_{c}(x)$ and $S_{s}(x)$ are the $c$ and $s$-quark propagators,
respectively. Explicit expressions of these propagators are presented in
the Appendix (see, also Ref.\ \cite{Agaev:2020zad}). In Eq.\ (\ref{eq:QCDside}),
we have also used the notation
\begin{equation}
\widetilde{S}_{c(s)}(x)=CS_{c(s)}^{T}(x)C.  \label{eq:Prop}
\end{equation}

\begin{figure}[h]
\includegraphics[width=8.5cm]{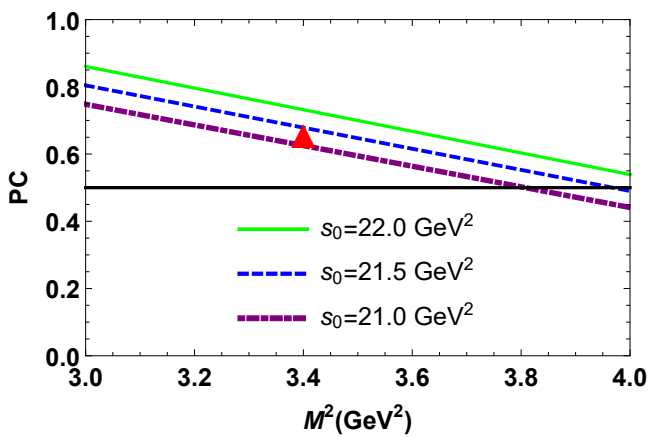}
\caption{Pole contribution as a function of the Borel parameter $M^{2}$ at
various $s_{0}$. The horizontal black line limits a region $\mathrm{PC}=0.5$%
. The red triangle fixes the point, where the mass $m$ of the tetraquark $X$
has effectively been extracted. }
\label{fig:PC}
\end{figure}
The correlation function $\Pi ^{\mathrm{OPE}}(p)$ should be computed in the
operator product expansion ($\mathrm{OPE}$) with some accuracy. The $\Pi ^{%
\mathrm{OPE}}(p)$ has also a trivial structure $\sim \mathrm{I}$ and is
characterized by an amplitude $\Pi ^{\mathrm{OPE}}(p^{2})$. Having equated
the invariant amplitudes $\Pi ^{\mathrm{Phys}}(p^{2})$ and $\Pi ^{\mathrm{OPE%
}}(p^{2})$, one gets the master QCD sum rule equality. Afterwards, one needs
to suppress contributions of higher resonances and continuum states by
applying the Borel transformation. The assumption about quark-hadron duality
allows one to subtract these suppressed terms from the obtained expression.
After these operations, the sum rule equality starts to depend on the Borel $%
M^{2}$ and continuum threshold $s_{0}$ parameters.

The Borel transformation of $\Pi ^{\mathrm{Phys}}(p^{2})$ is a simple
function, whereas for $\Pi ^{\mathrm{OPE}}(p^{2})$ we get a complicated
formula
\begin{equation}
\Pi (M^{2},s_{0})=\int_{4\mathcal{M}^{2}}^{s_{0}}ds\rho ^{\mathrm{OPE}%
}(s)e^{-s/M^{2}}+\Pi (M^{2}),  \label{eq:InvAmp}
\end{equation}%
where $\mathcal{M}=m_{c}+m_{s}$. In numerical computations we set $%
m_{s}^{2}=0$, but include into analysis terms proportional to $m_{s}$. The
two-point spectral density $\rho ^{\mathrm{OPE}}(s)$ is calculated as an
imaginary part of the correlation function. The second term $\Pi (M^{2})$
includes nonperturbative contributions extracted directly from $\Pi ^{%
\mathrm{OPE}}(p)$. The correlator $\Pi (M^{2},s_{0})$ is computed by taking
into account nonperturbative terms up to dimension $10$. Explicit expression
of $\Pi (M^{2},s_{0})$ is written down in the Appendix.

The sum rules for $m$ and $f$ are expressed via the invariant amplitude $\Pi
(M^{2},s_{0})$,
\begin{equation}
m^{2}=\frac{\Pi ^{\prime }(M^{2},s_{0})}{\Pi (M^{2},s_{0})},  \label{eq:Mass}
\end{equation}%
and
\begin{equation}
f^{2}=\frac{e^{m^{2}/M^{2}}}{m^{2}}\Pi (M^{2},s_{0}),  \label{eq:Coupling}
\end{equation}%
where $\Pi ^{\prime }(M^{2},s_{0})=d\Pi (M^{2},s_{0})/d(-1/M^{2})$.

To carry out the numerical computations in accordance with Eqs.\ (\ref{eq:Mass})
and (\ref{eq:Coupling}), we have to fix values of different vacuum
condensates. The reason is that the sum rules for $m^{2}$ and $f^{2}$
through $\Pi (M^{2},s_{0})$ depend on the vacuum expectation values of
quark, gluon and mixed operators. The vacuum condensates, that enter to the
sum rules Eqs.\ (\ref{eq:Mass}) and (\ref{eq:Coupling}), are universal
quantities obtained from analysis of various hadronic processes \cite%
{Shifman:1978bx,Shifman:1978by,Ioffe:1981kw,Ioffe:2005ym,Narison:2015nxh}%
\begin{eqnarray}
&&\langle \overline{q}q\rangle =-(0.24\pm 0.01)^{3}~\mathrm{GeV}^{3},\
\langle \overline{s}s\rangle =(0.8\pm 0.1)\langle \overline{q}q\rangle ,
\notag \\
&&\langle \overline{s}g_{s}\sigma Gs\rangle =m_{0}^{2}\langle \overline{s}%
s\rangle ,\ m_{0}^{2}=(0.8\pm 0.1)~\mathrm{GeV}^{2},\   \notag \\
&&\langle \frac{\alpha _{s}G^{2}}{\pi }\rangle =(0.012\pm 0.004)~\mathrm{GeV}%
^{4},  \notag \\
&&\langle g_{s}^{3}G^{3}\rangle =(0.57\pm 0.29)~\mathrm{GeV}^{6},  \notag \\
&&m_{c}=(1.27\pm 0.02)~\mathrm{GeV},\ \ \ m_{s}=93_{-5}^{+11}~\mathrm{MeV}.
\label{eq:Parameters}
\end{eqnarray}%
It is seen that the vacuum condensate of the strange quark differs from the  $%
\langle 0|\overline{q}q|0\rangle $ \cite{Ioffe:1981kw}. The mixed
condensates $\langle \overline{q}g_{s}\sigma Gq\rangle $ and $\langle
\overline{s}g_{s}\sigma Gs\rangle $ are expressed in terms of the
corresponding quark condensates and the parameter $m_{0}^{2}$. The numerical
value of the latter was extracted from analysis of baryonic resonances in
Ref.\ \cite{Ioffe:2005ym}. For the gluon condensate $\langle
g^{3}G^{3}\rangle $, we use the estimate given in Ref.\ \cite%
{Narison:2015nxh}. This list also contains the masses of $c$ and $s$ quarks
in the $\overline{\mathrm{MS}}$-scheme from Ref.\ \cite{PDG:2022}.

\begin{figure}[h]
\includegraphics[width=8.5cm]{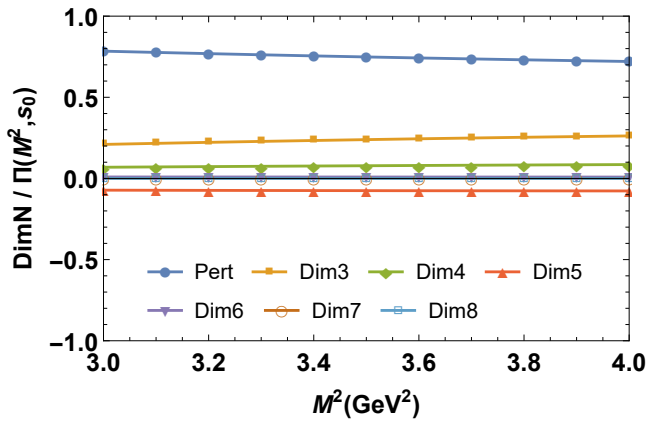}
\caption{Different contributions to $\Pi (M^{2},s_{0})$ normalized to $1$ as
functions of the Borel parameter $M^2$. All lines in this figure have been
calculated at $s_0=21.5~\mathrm{GeV}^2$. }
\label{fig:Convergence}
\end{figure}

Predictions for $m$ and $f$ extracted from the sum rules depend also on the
Borel and continuum subtraction parameters $M^{2}$ and $s_{0}$. In general,
physical quantities should not contain residual effects connected with the
choice of $M^{2}$. But in a real situation $m$ and $f$ \ bear imprints of
operations fulfilled to isolate contribution of the ground-state particle to
sum rules. A way to solve this problem is using some prescriptions to
minimize the unwanted effects. To this end, in the sum rule analysis, the choice
of a working window for the Borel parameter $M^{2}$ is restricted by the
dominance of the pole contribution ($\mathrm{PC}$) and convergence of $%
\mathrm{OPE}$. To quantify these constraints, it is convenient to introduce
the expressions
\begin{equation}
\mathrm{PC}=\frac{\Pi (M^{2},s_{0})}{\Pi (M^{2},\infty )},  \label{eqPC}
\end{equation}%
and
\begin{equation}
R(M^{2})=\frac{\Pi ^{\mathrm{DimN}}(M^{2},s_{0})}{\Pi (M^{2},s_{0})}.
\label{eq:Convergence}
\end{equation}%
First of them is a measure of the pole contribution and necessary to find
the higher border of the $M^{2}$ region. In Eq.\ (\ref{eq:Convergence}) $\Pi ^{%
\mathrm{DimN}}(M^{2},s_{0})$ indicates the  last three terms in $\mathrm{OPE}$ of
$\Pi (M^{2},s_{0})$, i.e., $\mathrm{DimN=Dim(8+9+10)}$. We use $R(M^{2})$ to
estimate the convergence of $\mathrm{OPE}$ and fix a lower limit of $M^{2}$.

In working regions of $M^{2}$ and $s_{0}$ the perturbative contribution to
the correlation function $\Pi (M^{2},s_{0})$ has to be larger than the ones due
to nonperturbative terms. Besides, the window for $M^{2}$ should generate
stable predictions for the extracted physical quantities. Performed analysis
demonstrates that windows for $M^{2}$ and $s_{0}$, which satisfy these
constraints,  are
\begin{equation}
M^{2}\in \lbrack 3,4]~\mathrm{GeV}^{2},\ s_{0}\in \lbrack 21,22]~\mathrm{GeV}%
^{2}.  \label{eq:Regions}
\end{equation}%
Indeed, in the regions Eq.\ (\ref{eq:Regions}) the pole contribution varies
on average within the interval
\begin{equation}
0.80\geq \mathrm{PC}\geq 0.49.  \label{eq:Polelimits}
\end{equation}%
In Fig.\ \ref{fig:PC}, the $\mathrm{PC}$ is drawn as a function of the Borel
parameter at various $s_{0}$. It is seen  that except for a small domain $%
M^{2}>2.8~\mathrm{GeV}^{2}$ at $s_{0}=21~\mathrm{GeV}^{2}$ the dominance of
the pole contribution, i.e., the constraint $\mathrm{PC}\geq 0.5$ is
fulfilled for all values of the parameters $M^{2}$ and $s_{0}$.

In Fig.\ \ref{fig:Convergence}, we demonstrate the dependence on $M^{2}$ of  the
perturbative and different nonperturbative contributions to $\Pi
(M^{2},s_{0})$. It is evident  that the perturbative term is considerably
larger than the  nonperturbative contributions, and constitutes $80\%$ of $\Pi
(M^{2},s_{0})$ at $M^{2}=3~\mathrm{GeV}^{2}$. This figure confirms also
convergence of the $\mathrm{OPE}$, which implies that the contributions of the
nonperturbative terms reduce by increasing the dimensions of the
corresponding operators. The $\mathrm{Dim3}$ term numerically exceeds
the contributions of other nonperturbative operators, whereas $\mathrm{Dim9}$
and $\mathrm{Dim10}$ terms are very small and not shown in the plot. The
quantity $R(M^{2})$ at $M^{2}=3~\mathrm{GeV}^{2}$ is less than $0.01$, which
proves numerically the convergence of the $\mathrm{OPE}$ and correctness of the
lower value of $M^{2}$.

\begin{widetext}

\begin{figure}[h!]
\begin{center}%%
\includegraphics[totalheight=6cm,width=8cm]{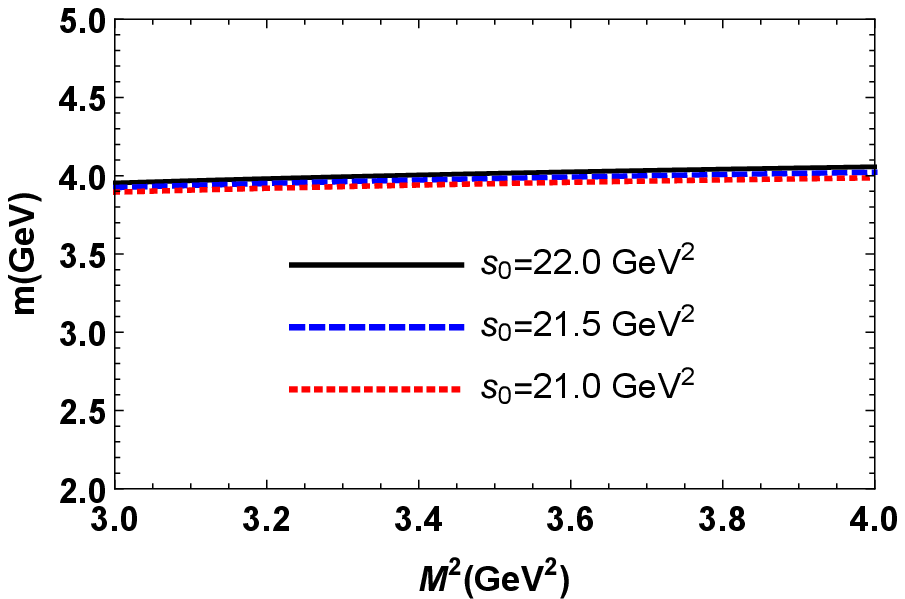}%%%
\includegraphics[totalheight=6cm,width=8cm]{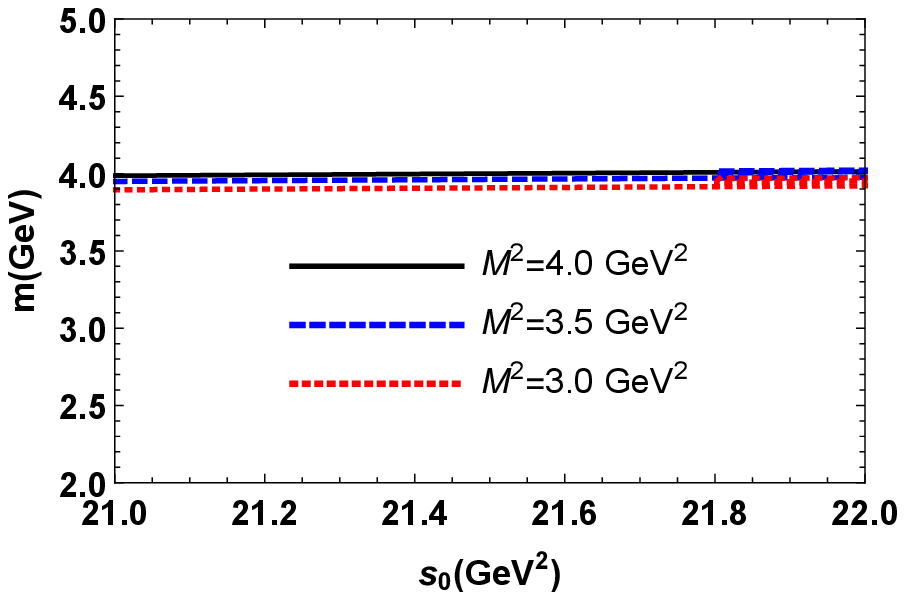}
\end{center}%%
\caption{Mass $m$ of the tetraquark $X$ as a function of the Borel $%%
M^{2}$
(left), and the continuum threshold
$s_0$ parameters
(right)}.%
\label{fig:Mass}
\end{figure}

\end{widetext}

The residual dependences of the mass $m$ of the tetraquark $X$ on the Borel
and continuum subtraction parameters $M^{2}$ and $s_{0}$ are shown in Fig.\ %
\ref{fig:Mass}. It is seen that the window for $M^{2}$, where parameters of
$X$ are extracted, leads to approximately stable predictions for $m$. At the
same time, one observes some variations of  $m$ against the Borel parameter $%
M^{2} $. This effect allows us to estimate the uncertainties of the sum rule
predictions. Variation of the continuum threshold parameter $s_{0}$ is
another source of the theoretical ambiguities. The region for $s_{0}$ has to
meet the constraints coming from the dominance of $\mathrm{PC}$ and convergence of the $%
\mathrm{OPE}$. The parameter $\sqrt{s_{0}}$ bears also information on the mass
$m^{\ast }$ of the first radial excitation of the tetraquark $X$, and should
obey $\sqrt{s_{0}}\leq m^{\ast }$.

The results for the mass $m$ and coupling $f$ are evaluated as mean values
of these quantities calculated in the working regions (\ref{eq:Regions}):
\begin{eqnarray}
m &=&(3976\pm 85)~\mathrm{MeV},  \notag \\
f &=&(7.3\pm 0.8)\times 10^{-3}~\mathrm{GeV}^{4}.  \label{eq:Result1}
\end{eqnarray}%
The mass and coupling written down in Eq.\ (\ref{eq:Result1}) effectively
correspond to the sum rule predictions at $M^{2}=3.4~\mathrm{GeV}^{2}$ and $%
s_{0}=21.5~\mathrm{GeV}^{2}$ shown in Fig.\ \ref{fig:PC} by the red
triangle. This point is located approximately at the middle of the working
regions, where the pole contribution is $\mathrm{PC}\approx 0.64$. This fact
and other details discussed above guarantees the ground-state nature of $X$
and credibility of the final results. An estimate for the mass of the
excited tetraquark $m^{\ast }\geq (m+650)~\mathrm{MeV}$ stemming from Eqs.\ (%
\ref{eq:Regions}) and (\ref{eq:Result1}) is also  reasonable for
the double-heavy tetraquarks.

%%%%%%%%%%%%%%%%%%%%%%%%%%%%%%%%%%%%%%%%%%%%%%%%%%%%%%%%%%%%%%%%%%%%%%

\section{Decay $X\rightarrow D_{s}^{+}D_{s}^{-}$}

\label{sec:Decay1}
%%%%%%%%%%%%%%%%%%%%%%%%%%%%%%%%%%%%%%%%%%%%%%%%%%%%%%%%%%%%%%%%%%%%%
The spectroscopic parameters of the tetraquark $X$ form a basis to determine
its kinematically allowed decay channels. Because $X(3960)$ was observed in
the $D_{s}^{+}D_{s}^{-}$ invariant mass distribution, we treat the decay $%
X\rightarrow D_{s}^{+}D_{s}^{-}$ as a dominant mode of $X$. The two-meson
threshold for this process $\approx 3937~\mathrm{MeV}$ is below the mass of $%
X$. Other decay channels that should be considered in this paper are $%
X\rightarrow \eta _{c}\eta ^{\prime }$ and $X\rightarrow \eta _{c}\eta $.
The kinematical limits for realization of these processes do not exceed $%
\approx 3941~\mathrm{MeV}$ which is less than $m$ as well. It is easy to see
also that decays of the scalar tetraquark with spin-parities $J^{\mathrm{PC}%
}=0^{++}$ to two pseudoscalar mesons with $J^{\mathrm{PC}}=0^{-+}$ preserves
the spin and quantum numbers $\mathrm{P}$ and $\mathrm{C}$ of the initial
state $X$.

The partial width of the decay $X\rightarrow D_{s}^{+}D_{s}^{-}$ is
determined by a coupling $G$ that describes the strong interaction at the vertex
$XD_{s}^{+}D_{s}^{-}$. Apart from $G$, it depends also on the  masses and decay
constants of the initial and final particles. The mass and coupling of $X$
have been calculated in the present article, whereas physical parameters of
the mesons $D_{s}^{+}$ and $D_{s}^{-}$ are known from other sources.
Therefore, the only physical quantity to be found here is the strong coupling $G$%
.

To evaluate $G$, we use the QCD three-point sum rule method, and start our
analysis from the correlation function
\begin{eqnarray}
&&\Pi (p,p^{\prime })=i^{2}\int d^{4}xd^{4}ye^{i(p^{\prime }y-px)}\langle 0|%
\mathcal{T}\{J^{D_{s}^{+}}(y)  \notag \\
&&\times J^{D_{s}^{-}}(0)J^{\dagger }(x)\}|0\rangle,  \label{eq:CF2}
\end{eqnarray}%
where $J(x)$,$\ J^{D_{s}^{+}}(y)$ and $J^{D_{s}^{-}}(0)$ are the
interpolating currents for the tetraquark $X$, and the pseudoscalar mesons $%
D_{s}^{+}$\ and $D_{s}^{-}$, respectively. The four-momenta of $X$ and $%
D_{s}^{+}$ are denoted by $p$ and $p^{\prime }$, whereas the momentum of the
meson $D_{s}^{-}$ is equal to $q=p-p^{\prime }$. The current $J(x)$ is given
by Eq.\ (\ref{eq:C1}), whereas for the mesons, we use the following currents:
\begin{eqnarray}
\ J^{D_{s}^{+}}(x) &=&\overline{s}_{j}(x)i\gamma _{5}c_{j}(x),  \notag \\
J^{D_{s}^{-}}(x) &=&\overline{c}_{i}(x)i\gamma _{5}s_{i}(x),  \label{eq:C2}
\end{eqnarray}%
with $i$ and $j$ being the color indices.

To continue our study of the strong coupling $G$, we follow usual recipes of
the sum rule method and compute the correlation function $\Pi (p,p^{\prime
}) $. To this end, we employ the physical parameters of the tetraquark and
mesons participating in this process. The correlator $\Pi (p,p^{\prime })$
found by this way constitutes the phenomenological side $\Pi ^{\mathrm{Phys}%
}(p,p^{\prime })$ of the sum rule. It is not difficult to see  that
\begin{eqnarray}
&&\Pi ^{\mathrm{Phys}}(p,p^{\prime })=\frac{\langle
0|J^{D_{s}^{+}}|D_{s}^{+}(p^{\prime })\rangle \langle
0|J^{D_{s}^{-}}|D_{s}^{-}(q)\rangle }{(p^{2}-m^{2})(p^{\prime
2}-m_{D_{s}}^{2})}  \notag \\
&&\times \frac{\langle D_{s}^{-}(q)D_{s}^{+}(p^{\prime })|X(p)\rangle
\langle X(p)|J^{\dagger }|0\rangle }{(q^{2}-m_{D_{s}}^{2})}+\cdots ,  \notag
\\
&&  \label{eq:CF3}
\end{eqnarray}%
where $m_{D_{s}}$ is the mass of the mesons $D_{s}^{\pm }$. To derive Eq.\ (%
\ref{eq:CF3}), we isolate the contribution of the ground-state particles
from ones due to higher resonances and continuum states. In Eq.\ (\ref%
{eq:CF3}) the ground-state term is presented explicitly, whereas the dots
stand for the other contributions.

The function $\Pi ^{\mathrm{Phys}}(p,p^{\prime })$ can be modified by
employing the matrix elements of the mesons $D_{s}^{\pm }$
\begin{equation}
\langle 0|J^{D_{s}^{\pm }}|D_{s}^{\pm }\rangle =\frac{m_{D_{s}}^{2}f_{D_{s}}%
}{m_{c}+m_{s}},\   \label{eq:Mel2}
\end{equation}%
with $f_{D_{s}}$ being their decay constants. The vertex $%
XD_{s}^{+}D_{s}^{-} $ is modeled as%
\begin{equation}
\langle D_{s}^{-}(q)D_{s}^{+}(p^{\prime })|X(p)\rangle =G(q^{2})p\cdot
p^{\prime }.  \label{eq:Ver1}
\end{equation}%
Using these matrix elements, one can easily find a new expression for $\Pi ^{%
\mathrm{Phys}}(p,p^{\prime })$:
\begin{eqnarray}
&&\Pi ^{\mathrm{Phys}}(p,p^{\prime })=G(q^{2})\frac{%
m_{D_{s}}^{4}f_{D_{s}}^{2}fm}{(m_{c}+m_{s})^{2}(p^{2}-m^{2})}  \notag \\
&&\times \frac{1}{(p^{\prime 2}-m_{D_{s}}^{2})(q^{2}-m_{D_{s}}^{2})}\frac{%
m^{2}+m_{D_{s}}^{2}-q^{2}}{2}+\cdots.  \notag \\
&&  \label{eq:Phys2}
\end{eqnarray}%
The double Borel transformation of the correlation function $\Pi ^{\mathrm{%
Phys}}(p,p^{\prime })$ over variables $p^{2}$ and $p^{\prime 2}$ is given by
the formula
\begin{eqnarray}
&&\mathcal{B}\Pi ^{\mathrm{Phys}}(p,p^{\prime })=G(q^{2})\frac{%
m_{D_{s}}^{4}f_{D_{s}}^{2}fm}{(m_{c}+m_{s})^{2}(q^{2}-m_{D_{s}}^{2})}%
e^{-m^{2}/M_{1}^{2}}  \notag \\
&&\times e^{-m_{D_{s}}^{2}/M_{2}^{2}}\frac{m^{2}+m_{D_{s}}^{2}-q^{2}}{2}%
+\cdots .  \label{eq:BTr}
\end{eqnarray}%
The correlator $\Pi ^{\mathrm{Phys}}(p,p^{\prime })$ and its Borel
transformation have a simple Lorentz structure which is proportional to $%
\mathrm{I}$. As a result, the relevant invariant amplitude $\Pi ^{\mathrm{%
Phys}}(p^{2},p^{\prime 2},q^{2})$ is determined by the whole expression
written down in Eq.\ (\ref{eq:Phys2}).

To derive the QCD side of the three-point sum rule, we express $\Pi
(p,p^{\prime })$ in terms of the quark propagators, and get
\begin{eqnarray}
&&\Pi ^{\mathrm{OPE}}(p,p^{\prime })=\int d^{4}xd^{4}ye^{i(p^{\prime
}y-px)}\epsilon \widetilde{\epsilon }  \notag \\
&&\times \mathrm{Tr}\left[ \gamma _{5}\widetilde{S}_{c}^{ib}(y-x)\gamma _{5}%
\widetilde{S}_{s}^{ni}(x-y)\gamma _{5}S_{c}^{mj}(x)\right.  \notag \\
&&\left. \times \gamma _{5}S_{s}^{jc}(-x)\right],  \label{eq:CF4}
\end{eqnarray}

The correlator $\Pi ^{\mathrm{OPE}}(p,p^{\prime })$ is computed by taking
into account the  nonperturbative contributions up to dimension $6$. This
function contains the same trivial Lorentz structure as $\Pi ^{\mathrm{Phys}%
}(p,p^{\prime })$. Having denoted by $\Pi ^{\mathrm{OPE}}(p^{2},p^{\prime
2},q^{2})$ the corresponding invariant amplitude, equated the double Borel
transformations $\mathcal{B}\Pi ^{\mathrm{OPE}}(p^{2},p^{\prime 2},q^{2})$
and $\mathcal{B}\Pi ^{\mathrm{Phys}}(p^{2},p^{\prime 2},q^{2})$, and
performed continuum subtraction, we find the sum rule for the strong
coupling $G(q^{2})$.

The amplitude $\Pi ^{\mathrm{OPE}}(p^{2},p^{\prime 2},q^{2})$ after the
Borel transformation and continuum subtraction procedures can be expressed
using the spectral density $\rho (s,s^{\prime },q^{2})$ which is
proportional to a relevant imaginary part of $\Pi ^{\mathrm{OPE}%
}(p,p^{\prime })$
\begin{eqnarray}
&&\Pi (\mathbf{M}^{2},\mathbf{s}_{0},q^{2})=\int_{4\mathcal{M}%
^{2}}^{s_{0}}ds\int_{\mathcal{M}^{2}}^{s_{0}^{\prime }}ds^{\prime }\rho
(s,s^{\prime },q^{2})  \notag \\
&&\times e^{-s/M_{1}^{2}}e^{-s^{\prime }/M_{2}^{2}}.  \label{eq:SCoupl}
\end{eqnarray}%
The Borel and continuum threshold parameters are denoted in Eq.\ (\ref%
{eq:SCoupl}) by $\mathbf{M}^{2}=(M_{1}^{2},\ M_{2}^{2})$ and $\mathbf{s}%
_{0}=(s_{0},\ s_{0}^{\prime })$, respectively. Then, the sum rule for $%
G(q^{2})$ reads
\begin{eqnarray}
G(q^{2}) &=&\frac{2(m_{c}+m_{s})^{2}}{m_{D_{s}}^{4}f_{D_{s}}^{2}fm}\frac{%
q^{2}-m_{D_{s}}^{2}}{m^{2}+m_{D_{s}}^{2}-q^{2}}  \notag \\
&&\times e^{m^{2}/M_{1}^{2}}e^{m_{D_{s}}^{2}/M_{2}^{2}}\Pi (\mathbf{M}^{2},%
\mathbf{s}_{0},q^{2}).  \label{eq:SRCoup}
\end{eqnarray}%
The coupling $G(q^{2})$ is also a function of the Borel and continuum
threshold parameters, which, for the sake of simplicity, are not shown in
Eq.\ (\ref{eq:SRCoup}). In what follows, we introduce a variable $%
Q^{2}=-q^{2}$ and label the obtained function $G(Q^{2})$.

Equation (\ref{eq:SRCoup}) contains the spectroscopic parameters of the
tetraquark $X$, and the masses and decay constants of the mesons $D_{s}^{\pm
}$. These parameters are input information for our numerical computations:
Their values are collected in Table\ \ref{tab:Param}, which contains also
parameters of the mesons $\eta _{c}$, $\eta ^{\prime }$ and $\eta $
appearing at final stages of the other processes. For the masses of the
mesons and decay constant $f_{D_{s}}$, we use information from Ref.\ \cite%
{PDG:2022}. As the decay constant of the meson $\eta _{c}$, we employ sum
rule's prediction from Ref.\ \cite{Colangelo:1992cx}.

For numerical calculations of $G(Q^{2})$ one has to fix working windows for
the Borel and continuum subtraction parameters $\mathbf{M}^{2}$ and $\mathbf{%
s}_{0}$. The constraints imposed on $\mathbf{M}^{2}$ and $\mathbf{s}_{0}$
are usual for sum rule calculations: They have been discussed and explained
in the section \ref{sec:Mass}. The regions for $M_{1}^{2}$ and $s_{0}$, that
correspond to the $X$ channel, are chosen as in Eq.\ (\ref{eq:Regions}). The
parameters $(M_{2}^{2},\ s_{0}^{\prime })$ for the $D_{s}^{+}$ meson channel
are varied within limits
\begin{equation}
M_{2}^{2}\in \lbrack 2.5,3.5]~\mathrm{GeV}^{2},\ s_{0}^{\prime }\in \lbrack
5,6]~\mathrm{GeV}^{2}.  \label{eq:Wind3}
\end{equation}%
The windows Eq.\ (\ref{eq:Wind3}) are well correlated with the $D_{s}^{+}$
meson's mass. In fact, $\sqrt{s_{0}^{\prime }}\approx (m_{D_{s}}+0.35)\
\mathrm{GeV}$ is a typical choice for mesons with experimentally measured
masses. The Borel parameter $M_{2}$ is also comparable with the mass of the $%
D_{s}^{+}$ meson. The regions Eq.\ (\ref{eq:Wind3}) are numerically very
close to ones given in our article Ref.\ \cite{Agaev:2022ast} for the $%
D^{\ast +}$ channel in the decay $M_{cc}^{+}\rightarrow D^{0}D^{\ast +}$.
Nevertheless, a decisive factor in choice of $(M_{2}^{2},\ s_{0}^{\prime })$
is fulfillment of the sum rule constraints.

\begin{figure}[h]
\includegraphics[width=8.5cm]{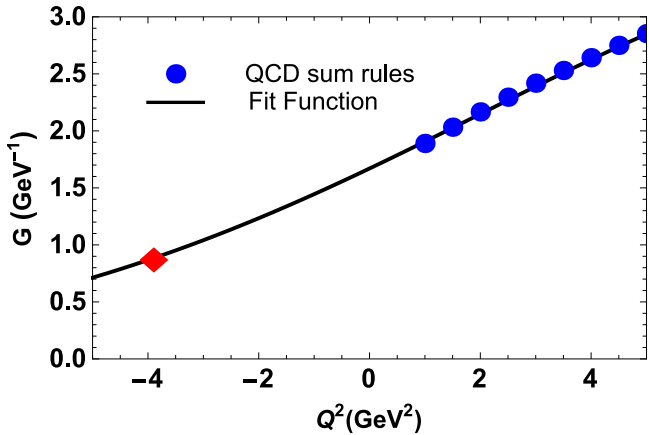}
\caption{The sum rule predictions and fit function for the strong coupling $%
G(Q^{2})$. The point $Q^{2}=-m_{D_s}^{2}$ is shown by the red diamond. }
\label{fig:Fit}
\end{figure}

\begin{figure}[h]
\includegraphics[width=8.8cm]{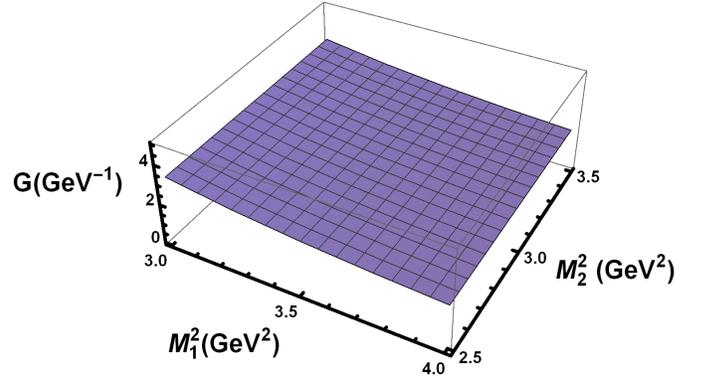}
\caption{ The strong coupling $G=G(3~\mathrm{GeV}^{2})$ as a function of the
Borel parameters $M_{1}^{2}$ and $M_{2}^{2}$ at $s_{0}=21.5~\mathrm{GeV}^{2}$
and $s_{0}^{\prime }=5.5~\mathrm{GeV}^{2}$.}
\label{fig:StrongC}
\end{figure}

Thus, we calculate $G(Q^{2})$ at fixed $Q^{2}=1-5~\mathrm{GeV}^{2}$ and
depict obtained the results in Fig.\ \ref{fig:Fit}. Let us emphasize that
the constraints imposed on parameters $\mathbf{M}^{2}$ and $\mathbf{s}_{0}$ by
the sum rule analysis are satisfied at each $Q^{2}$. For instance, in Fig.\ %
\ref{fig:StrongC} the coupling $G(Q^{2})$ is plotted as a function of the
parameters $M_{1}^{2}$ and $M_{2}^{2}$ at $Q^{2}=3~\mathrm{GeV}^{2}$ and
middle of the $s_{0}$ and $s_{0}^{\prime }$ regions. Variations of $G(3~%
\mathrm{GeV}^{2})$ while changing $M_{1}^{2}$ and $M_{2}^{2}$ in explored
regions stay within acceptable limits and do not exceed $\pm 25\%$ of the
central value. Numerically, we find
\begin{equation}
G(3\ \mathrm{GeV}^{2})=(2.53\pm 0.62)~\mathrm{GeV}^{-1}.
\end{equation}

The partial width of the process $X\rightarrow D_{s}^{+}D_{s}^{-}$ should be
calculated in terms of the strong coupling $G(-m_{D_{s}}^{2})$ which is
defined at the mass shell $q^{2}=m_{D_{s}}^{2}$ of the meson $D_{s}^{-}$.
But the region $Q^{2}<0$ is not accessible for the sum rule analysis. To
solve this problem, it is convenient to introduce a fit function $\mathcal{G}%
_{1}(Q^{2})$ which for the momenta $Q^{2}>0$ is consistent with predictions
of the sum rule computations, but can be extrapolated to the region $Q^{2}<0$%
. For these purposes, we apply the function $\mathcal{G}_{i}(Q^{2}),\
i=0,1,2 $
\begin{equation}
\mathcal{G}_{i}(Q^{2})=\mathcal{G}_{i}^{0}\mathrm{\exp }\left[ c_{i}^{1}%
\frac{Q^{2}}{m^{2}}+c_{i}^{2}\left( \frac{Q^{2}}{m^{2}}\right) ^{2}\right] ,
\label{eq:FitF}
\end{equation}%
where $\mathcal{G}_{i}^{0}$, $c_{i}^{1}$ and $c_{i}^{2}$ are parameters,
which will be extracted from fitting procedures. Numerical calculations
demonstrate that $\mathcal{G}_{0}^{0}=1.67~\mathrm{GeV}^{-1}$, $%
c_{0}^{1}=2.19$, and $c_{0}^{2}=-1.59$ generate a nice agreement with the
sum rule's data shown in Fig.\ \ref{fig:Fit}.

At the mass shell $q^{2}=m_{D_{s}}^{2}$ this function predicts
\begin{equation}
G\equiv \mathcal{G}_{0}(-m_{D_{s}}^{2})=(8.9\pm 2.2)\times 10^{-1}\ \mathrm{%
GeV}^{-1}.  \label{eq:Coupl1}
\end{equation}%
The width of the process $X\rightarrow D_{s}^{+}D_{s}^{-}$ is determined by
the following formula%
\begin{equation}
\Gamma \left[ X\rightarrow D_{s}^{+}D_{s}^{-}\right] =G^{2}\frac{%
m_{D_{s}}^{2}\lambda }{8\pi }\left( 1+\frac{\lambda ^{2}}{m_{D_{s}}^{2}}%
\right) ,  \label{eq:PartDW}
\end{equation}%
where $\lambda =\lambda \left( m,m_{D_{s}},m_{D_{s}}\right) $ and
\begin{eqnarray}
\lambda \left( a,b,c\right) &=&\frac{1}{2a}\left[ a^{4}+b^{4}+c^{4}\right.
\notag \\
&&\left. -2\left( a^{2}b^{2}+a^{2}c^{2}+b^{2}c^{2}\right) \right] ^{1/2}.
\end{eqnarray}%
Employing the coupling from Eq.\ (\ref{eq:Coupl1}), it is easy to find
partial width of the process $X\rightarrow D_{s}^{+}D_{s}^{-}$%
\begin{equation}
\Gamma \left[ X\rightarrow D_{s}^{+}D_{s}^{-}\right] =(34.0\pm 11.9)~\mathrm{%
MeV}.  \label{eq:DW1Numeric}
\end{equation}

\begin{table}[tbp]
\begin{tabular}{|c|c|}
\hline\hline
Quantity & Value (in $\mathrm{MeV}$ units) \\ \hline
$m_{D_s}$ & $1969.0\pm 1.4$ \\
$m_{\eta_{c}}$ & $2983.9 \pm 0.4$ \\
$m_{\eta^{\prime}}$ & $957.78 \pm 0.06$ \\
$m_{\eta}$ & $547.862 \pm 0.017$ \\
$f_{D_s}$ & $249.9 \pm 0.5$ \\
$f_{\eta_c}$ & $320 \pm 40$ \\ \hline\hline
\end{tabular}%
\caption{Masses and decay constants of the mesons $D_{s}^{\pm}$, $\protect%
\eta_c$, $\protect\eta^{\prime}$, and $\protect\eta$ which are employed in
numerical calculations.}
\label{tab:Param}
\end{table}

%%%%%%%%%%%%%%%%%%%%%%%%%%%%%%%%%%%%%%%%%%%%%%%%%%%%%%%%%%%%%%%%%%%%%%%%%%%%%%%%%%%

\section{Processes $X\rightarrow \protect\eta _{c}\protect\eta ^{\prime }$
and $X\rightarrow \protect\eta _{c}\protect\eta $}

\label{sec:Decays2}
%%%%%%%%%%%%%%%%%%%%%%%%%%%%%%%%%%%%%%%%%%%%%%%%%%%%%%%%%%%%%%%%%%%%%%%%%%%%%%%%%%%

The processes $X\rightarrow \eta _{c}\eta ^{\prime }$ and $X\rightarrow \eta
_{c}\eta $, in general, can be studied by a manner described above. But, it
is well known that due to $U(1)$ anomaly there is a mixing in the system of $%
\eta -\eta ^{\prime }$ mesons \cite{Feldmann:1998vh}. This phenomenon leads
to some subtleties in the choice of the  interpolating currents for these
particles. The $\eta -\eta ^{\prime }$ mixing can be described in the
framework of different approaches: The physical particles $\eta $ and $\eta
^{\prime }$ can be expressed using either the octet-singlet or quark-flavor
bases of the flavor $SU_{f}(3)$ group. It turns out that mixing of the
physical states, decay constants and higher twist distribution amplitudes in
the $\eta -\eta ^{\prime }$ system take simple forms in the quark-flavor basis $|\eta
_{q}\rangle =(\overline{u}u+\overline{d}d)/\sqrt{2}$ and $|\eta _{s}\rangle =%
\overline{s}s$  \cite%
{Feldmann:1998vh,Agaev:2014wna,Agaev:2015faa}. Therefore, for our purposes
it is convenient to describe the mesons $\eta $ and $\eta ^{\prime }$ in the
quark-flavor basis.

Then, the physical mesons $\eta $ and $\eta ^{\prime }$ are expressed using
the basic states $|\eta _{q}\rangle $ and $|\eta _{s}\rangle $%
\begin{equation}
\begin{pmatrix}
\eta \\
\eta ^{\prime }%
\end{pmatrix}%
=U(\varphi )%
\begin{pmatrix}
|\eta _{q}\rangle \\
|\eta _{s}\rangle%
\end{pmatrix}%
,  \label{eq:MixPhys}
\end{equation}%
where
\begin{equation}
U(\varphi )=%
\begin{pmatrix}
\cos \varphi & -\sin \varphi \\
\sin \varphi & \cos \varphi%
\end{pmatrix}%
,  \label{eq:Mixmat}
\end{equation}%
is the mixing matrix in $|\eta _{q}\rangle -|\eta _{s}\rangle $ basis with $%
\varphi $ being a mixing angle. This assumption on the state mixing implies
that the same pattern applies to relevant currents, decay constants and wave
functions as well.

In this context the interpolating currents for the mesons $\eta $ and $\eta
^{\prime }$ are given by the expressions%
\begin{eqnarray}
J^{\eta }(x) &=&-\sin \varphi \overline{s}_{j}(x)i\gamma _{5}s_{j}(x),
\notag \\
J^{\eta ^{\prime }}(x) &=&\cos \varphi \overline{s}_{j}(x)i\gamma
_{5}s_{j}(x),  \label{eq:EtaCurr}
\end{eqnarray}%
where $j$ is the color index. Let us emphasize that in Eq.\ (\ref{eq:EtaCurr}%
), we write down only $\overline{s}s$  component of the currents, which contribute to the
decays under analysis.

We begin our calculations from the decay $X\rightarrow \eta _{c}\eta
^{\prime }$. In this case, one should explore the correlation function
\begin{eqnarray}
\widetilde{\Pi }(p,p^{\prime }) &=&i^{2}\int d^{4}xd^{4}ye^{i(p^{\prime
}y-px)}\langle 0|\mathcal{T}\{J^{\eta _{c}}(y)  \notag \\
&&\times J^{\eta ^{\prime }}(0)J^{\dagger }(x)\}|0\rangle ,  \label{eq:CF2a}
\end{eqnarray}%
with $\ J^{\eta _{c}}(y)$ being the interpolating current of the meson $\eta
_{c}$
\begin{equation}
\ J^{\eta _{c}}(x)=\overline{c}_{i}(x)i\gamma _{5}c_{i}(x).
\label{eq:Curr5a}
\end{equation}
The ground-state contribution to the correlation function $\widetilde{\Pi }%
(p,p^{\prime })$ in terms of the involved particles' matrix elements has the
form
\begin{eqnarray}
&&\widetilde{\Pi }^{\mathrm{Phys}}(p,p^{\prime })=\frac{\langle 0|J^{\eta
_{c}}|\eta _{c}(p^{\prime })\rangle \langle 0|J^{\eta ^{\prime }}|\eta
^{\prime }(q)\rangle }{(p^{2}-m^{2})(p^{\prime 2}-m_{\eta _{c}}^{2})}  \notag
\\
&&\times \frac{\langle \eta ^{\prime }(q)\eta _{c}(p^{\prime })|X(p)\rangle
\langle X(p)|J^{\dagger }|0\rangle }{(q^{2}-m_{\eta ^{\prime }}^{2})}+\cdots
,  \notag \\
&&  \label{eq:CF5}
\end{eqnarray}%
where the dots indicate effects of higher resonances and continuum states.
The function $\widetilde{\Pi }^{\mathrm{Phys}}(p,p^{\prime })$ can be
simplified by invoking the matrix elements of the mesons $\eta _{c}$ and $%
\eta ^{\prime }$
\begin{eqnarray}
&&\langle 0|J^{\eta _{c}}|\eta _{c}\rangle =\frac{m_{\eta _{c}}^{2}\ f_{\eta
_{c}}}{2m_{c}},\   \notag \\
&&2m_{s}\langle \eta ^{\prime }|\overline{s}i\gamma _{5}s|0\rangle =h_{\eta
^{\prime }}^{s},  \label{eq:Mel3}
\end{eqnarray}%
where $m_{\eta _{c}}$ and $\ f_{\eta _{c}}$ are the mass and decay constant
of the $\eta _{c}$ meson. The twist-3 matrix element of the local operator $%
\overline{s}i\gamma _{5}s$ sandwiched between the meson $\eta ^{\prime }$
and vacuum states is denoted by $h_{\eta ^{\prime }}^{s}$ \cite%
{Agaev:2014wna}. \ The parameter $h_{\eta ^{\prime }}^{s}$ complies with
mixing effect, and we get%
\begin{equation}
h_{\eta ^{\prime }}^{s}=h_{s}\cos \varphi .  \label{eq:Mel4}
\end{equation}%
The parameter $h_{s}$ in Eq.\ (\ref{eq:Mel4}) can be defined theoretically
\cite{Agaev:2014wna}, but for our analysis it is enough to use phenomenological values of  $h_{s}$ and $\varphi $
\begin{eqnarray}
h_{s} &=&(0.087\pm 0.006)~\mathrm{GeV}^{3},  \notag \\
\varphi &=&39.3^{\circ }\pm 1.0^{\circ }.  \label{eq:MixParam}
\end{eqnarray}

The vertex $X\eta _{c}\eta ^{\prime }$ is chosen in the following form%
\begin{equation}
\langle \eta ^{\prime }(q)\eta _{c}(p^{\prime })|X(p)\rangle
=g_{1}(q^{2})p\cdot p^{\prime },
\end{equation}%
where $g_{1}$ is the strong coupling corresponding to the vertex $X\eta
_{c}\eta ^{\prime }$. Using these matrix elements, one can obtain a new
expression for $\Pi ^{\mathrm{Phys}}(p,p^{\prime })$
\begin{eqnarray}
&&\widetilde{\Pi }^{\mathrm{Phys}}(p,p^{\prime })=g_{1}(q^{2})\frac{%
fmm_{\eta _{c}}^{2}\ f_{\eta _{c}}h_{s}\cos ^{2}\varphi }{%
4m_{c}m_{s}{}(p^{2}-m^{2})}  \notag \\
&&\times \frac{1}{(p^{\prime 2}-m_{\eta _{c}}^{2})(q^{2}-m_{\eta ^{\prime
}}^{2})}\frac{m^{2}+m_{\eta _{c}}^{2}-q^{2}}{2}+\cdots .  \notag \\
&&
\end{eqnarray}%
The QCD side of the sum rule for $g_{1}(q^{2})$ is given by the formula
\begin{eqnarray}
&&\widetilde{\Pi }^{\mathrm{OPE}}(p,p^{\prime })=-\cos \varphi \int
d^{4}xd^{4}ye^{i(p^{\prime }y-px)}\epsilon \widetilde{\epsilon }  \notag \\
&&\times \mathrm{Tr}\left[ \gamma _{5}S_{c}^{ib}(y-x)\gamma _{5}\widetilde{S}%
_{s}^{jc}(-x)\gamma _{5}\widetilde{S}_{s}^{nj}(x)\right.  \notag \\
&&\left. \times \gamma _{5}S_{c}^{mi}(x-y)\right] ,  \label{eq:CF6}
\end{eqnarray}%
The sum rule for the coupling $g_{1}(q^{2})$ is derived using Borel
transformations of invariant amplitudes $\widetilde{\Pi }^{\mathrm{Phys}%
}(p^{2},p^{\prime 2},q^{2})$ and $\widetilde{\Pi }^{\mathrm{OPE}%
}(p^{2},p^{\prime 2},q^{2})$ and reads%
\begin{eqnarray}
g_{1}(q^{2}) &=&-\frac{8m_{c}m_{s}}{fmm_{\eta _{c}}^{2}\ f_{\eta
_{c}}h_{s}\cos \varphi }\frac{q^{2}-m_{\eta _{c}}^{2}}{m^{2}+m_{\eta
_{c}}^{2}-q^{2}}  \notag \\
&&\times e^{m^{2}/M_{1}^{2}}e^{m_{\eta _{c}}^{2}/M_{2}^{2}}\widetilde{\Pi }(%
\mathbf{M}^{2},\mathbf{s}_{0},q^{2}).  \label{eq:StCoup1}
\end{eqnarray}%
Here, $\widetilde{\Pi }(\mathbf{M}^{2},\mathbf{s}_{0},q^{2})$ is the Borel
transformed and subtracted amplitude $\widetilde{\Pi }^{\mathrm{OPE}%
}(p^{2},p^{\prime 2},q^{2})$.

The coupling $g_{1}(q^{2})$ is calculated using the following Borel and
continuum threshold parameters in the $\eta _{c}$ channel
\begin{equation}
M_{2}^{2}\in \lbrack 3,4]~\mathrm{GeV}^{2},\ s_{0}^{\prime }\in \lbrack
9.5,10.5]~\mathrm{GeV}^{2},  \label{eq.Wind4}
\end{equation}%
whereas as $M_{1}^{2}$ and $s_{0}$ for the $X$ channel, we employ Eq.\ (\ref%
{eq:Regions}). The strong coupling $g_{1}$ is defined at the mass shell of
the $\eta ^{\prime }$ meson. The fit function $\mathcal{G}_{1}(Q^{2})$ given
by Eq.\ (\ref{eq:FitF}) has the parameters $\mathcal{G}_{1}^{0}=0.26~\mathrm{%
GeV}^{-1}$, $c_{1}^{1}=4.72$, and $c_{1}^{2}=-3.52$. Relevant computations
yield%
\begin{equation}
g_{1}\equiv \mathcal{G}_{1}(-m_{\eta ^{\prime }}^{2})=(1.9\pm 0.3)\times
10^{-1}\ \mathrm{GeV}^{-1}.  \label{eq:g1}
\end{equation}

The partial width of this decay can be found by means of the formula Eq.\ (%
\ref{eq:PartDW}), in which one should make substitutions $G\rightarrow g_{1}$%
, $m_{D_{s}}^{2}\rightarrow m_{\eta _{c}}^{2}$ and $\lambda \left(
m,m_{D_{s}},m_{D_{s}}\right) \rightarrow \widetilde{\lambda }\left(
m,m_{\eta _{c}},m_{\eta ^{\prime }}\right) $. Then, for the process $%
X\rightarrow \eta _{c}\eta ^{\prime }$, we get
\begin{equation}
\Gamma \left[ X\rightarrow \eta _{c}\eta ^{\prime }\right] =(3.0\pm 0.7)~%
\mathrm{MeV}.  \label{eq:DW2Numeric}
\end{equation}

Analysis of the decay $X\rightarrow \eta _{c}\eta $ can be performed in a
similar way. Omitting further details, let us write down predictions
obtained for key quantities. Thus, the strong coupling $g_{2}$ at the vertex
$X\eta _{c}\eta $ is determined by the equality
\begin{equation}
g_{2}\equiv |\mathcal{G}_{2}(-m_{\eta }^{2})|=(1.4\pm 0.2)\times 10^{-1}\
\mathrm{GeV}^{-1},  \label{eq:g2}
\end{equation}%
where parameters of the fit function are $\mathcal{G}_{2}^{0}=-0.15~\mathrm{%
GeV}^{-1}$, $c_{2}^{1}=5.76$, and $c_{2}^{2}=-4.44$. The partial width of
the decay $X\rightarrow \eta _{c}\eta $ is%
\begin{equation}
\Gamma \left[ X\rightarrow \eta _{c}\eta \right] =(5.2\pm 1.1)~\mathrm{MeV}.
\label{eq:DW3}
\end{equation}

With this information at hands, it is not difficult to find the full width
of the scalar tetraquark $X$
\begin{equation}
\Gamma _{\mathrm{X}}=(42.2\pm 12.0)~\mathrm{MeV}.  \label{eq:FullW}
\end{equation}%
This estimate is in excellent agreement with the LHCb data.

%%%%%%%%%%%%%%%%%%%%%%%%%%%%%%%%%%%%%%%%%%%%%%%%%%%%%%%%%%%%%%%%%%%%%%%%%%

\section{Concluding notes}

\label{sec:Conclusion}
%%%%%%%%%%%%%%%%%%%%%%%%%%%%%%%%%%%%%%%%%%%%%%%%%%%%%%%%%%%%%%%%%%%%%%%%%%
In this article, we have calculated spectral parameters of the scalar
tetraquark $X$ in the framework of the QCD two-point sum rule method. We
evaluated also full width of $X$, by taking into account its decay modes $%
X\rightarrow D_{s}^{+}D_{s}^{-}$, $X\rightarrow \eta _{c}\eta ^{\prime }$
and $X\rightarrow \eta _{c}\eta $. Our result for the mass $m=(3976\pm 85)~%
\mathrm{MeV}$ of the tetraquark $X$ overshoots the corresponding LHCb datum,
but is compatible with $m_{\exp }$ provided one takes into account
the corresponding theoretical and experimental errors. Our prediction for the
full width $\Gamma _{\mathrm{X}}=(42.2\pm 12.0)~\mathrm{MeV}$ of $X$ is in
excellent agreement with $\Gamma _{\mathrm{exp}}$ from Eq.\ (\ref{eq:Data}).

In Ref.\ \cite{LHCb:2022vsv} the LHCb Collaboration assumed that the
resonance $X(3960)$ is composed of four $c\overline{c}s\overline{s}$ quarks.
This assumption is relied on theoretical predictions of Ref.\ \cite%
{Chen:2017dpy}, in which the authors used QCD sum rule method and different
interpolating currents to find the mass spectra of the diquark-antidiquark states $qc%
\overline{q}\overline{c}$ and $sc\overline{s}\overline{c}$ with $J^{\mathrm{%
PC}}=0^{++}$ and $2^{++}$. Some of the currents used there indeed led to
estimations which are comparable with $m_{\mathrm{exp} }$ if one includes
into consideration the ambiguities of the analysis.

In the context of the sum rule approach,  $X(3960)$ was modeled as a $%
D_{s}^{+}D_{s}^{-}$ molecule state \cite{Xin:2022bzt}, as well. In
accordance with this paper, the mass of such hadronic molecule is equal to $%
(3980\pm 100)~\mathrm{MeV}$ being in agreement with the LHCb data. It is
worth noting that in Refs.\ \cite{Chen:2017dpy,Xin:2022bzt} the authors did
not investigate quantitatively widths of the diquark-antidiquark or molecule
states considered there, which is crucial to draw a conclusion about the  inner
organization of $X(3960)$.

The results for $m$ and $\Gamma _{\mathrm{X}}$ obtained in the present
article allow us to consider $X(3960)$ as a candidate to a scalar
diquark-antidiquark exotic meson. At the same time, a molecule model for $%
X(3960)$ should be studied in a more detailed form: There is a necessity to
evaluate full width of a molecule state. Only after such comprehensive
analysis it is possible to make a choice between the  competing models.

\begin{widetext}

%%%%%%%%%%%%%%%%%%%%%%%%%%%%%%%%%%%%%%%%%%%%%%%%%%%%%%%%%%%%%%%%%%%%%%%%
\appendix*

\section{ The propagators $S_{q(Q)}(x)$ and the invariant amplitude $\Pi
(M^{2},s_{0})$}

\renewcommand{\theequation}{\Alph{section}.\arabic{equation}} \label{sec:App}
%%%%%%%%%%%%%%%%%%%%%%%%%%%%%%%%%%%%%%%%%%%%%%%%%%%%%%%%%%%%%%%%%%%%%%

In the current article, for the light quark propagator $S_{q}^{ab}(x)$, we
employ the following expression
\begin{eqnarray}
&&S_{q}^{ab}(x)=i\delta _{ab}\frac{\slashed x}{2\pi ^{2}x^{4}}-\delta _{ab}%
\frac{m_{q}}{4\pi ^{2}x^{2}}-\delta _{ab}\frac{\langle \overline{q}q\rangle
}{12}+i\delta _{ab}\frac{\slashed xm_{q}\langle \overline{q}q\rangle }{48}%
-\delta _{ab}\frac{x^{2}}{192}\langle \overline{q}g_{s}\sigma Gq\rangle
\notag \\
&&+i\delta _{ab}\frac{x^{2}\slashed xm_{q}}{1152}\langle \overline{q}%
g_{s}\sigma Gq\rangle -i\frac{g_{s}G_{ab}^{\alpha \beta }}{32\pi ^{2}x^{2}}%
\left[ \slashed x{\sigma _{\alpha \beta }+\sigma _{\alpha \beta }}\slashed x%
\right] -i\delta _{ab}\frac{x^{2}\slashed xg_{s}^{2}\langle \overline{q}%
q\rangle ^{2}}{7776}  \notag \\
&&-\delta _{ab}\frac{x^{4}\langle \overline{q}q\rangle \langle
g_{s}^{2}G^{2}\rangle }{27648}+\cdots.
\end{eqnarray}%
For the heavy quark $Q=c$, we use the propagator $S_{Q}^{ab}(x)$
\begin{eqnarray}
&&S_{Q}^{ab}(x)=i\int \frac{d^{4}k}{(2\pi )^{4}}e^{-ikx}\Bigg \{\frac{\delta
_{ab}\left( {\slashed k}+m_{Q}\right) }{k^{2}-m_{Q}^{2}}-\frac{%
g_{s}G_{ab}^{\alpha \beta }}{4}\frac{\sigma _{\alpha \beta }\left( {\slashed %
k}+m_{Q}\right) +\left( {\slashed k}+m_{Q}\right) \sigma _{\alpha \beta }}{%
(k^{2}-m_{Q}^{2})^{2}}  \notag \\
&&+\frac{g_{s}^{2}G^{2}}{12}\delta _{ab}m_{Q}\frac{k^{2}+m_{Q}{\slashed k}}{%
(k^{2}-m_{Q}^{2})^{4}}+\frac{g_{s}^{3}G^{3}}{48}\delta _{ab}\frac{\left( {%
\slashed k}+m_{Q}\right) }{(k^{2}-m_{Q}^{2})^{6}}\left[ {\slashed k}\left(
k^{2}-3m_{Q}^{2}\right) +2m_{Q}\left( 2k^{2}-m_{Q}^{2}\right) \right] \left(
{\slashed k}+m_{Q}\right) +\cdots \Bigg \}.  \notag \\
&&
\end{eqnarray}

Here, we have used the short-hand notations
\begin{equation}
G_{ab}^{\alpha \beta }\equiv G_{A}^{\alpha \beta }\lambda _{ab}^{A}/2,\ \
G^{2}=G_{\alpha \beta }^{A}G_{A}^{\alpha \beta },\ G^{3}=f^{ABC}G_{\alpha
\beta }^{A}G^{B\beta \delta }G_{\delta }^{C\alpha },
\end{equation}%
where $G_{A}^{\alpha \beta }$ is the gluon field strength tensor, $\lambda
^{A}$ and $f^{ABC}$ are the Gell-Mann matrices and structure constants of
the color group $SU_{c}(3)$, respectively. The indices $A,B,C$ run in the
range $1,2,\ldots 8$.

The invariant amplitude $\Pi (M^{2},s_{0})$ obtained after the Borel
transformation and subtraction procedures is given by Eq.\ (\ref{eq:InvAmp})%
\begin{equation*}
\Pi (M^{2},s_{0})=\int_{4\mathcal{M}^{2}}^{s_{0}}ds\rho ^{\mathrm{OPE}%
}(s)e^{-s/M^{2}}+\Pi (M^{2}),
\end{equation*}%
where the spectral density $\rho ^{\mathrm{OPE}}(s)$ and the function $\Pi
(M^{2})$ are determined by the expressions
\begin{equation}
\rho ^{\mathrm{OPE}}(s)=\rho ^{\mathrm{pert.}}(s)+\sum_{N=3}^{8}\rho ^{%
\mathrm{DimN}}(s),\ \ \Pi (M^{2})=\sum_{N=6}^{10}\Pi ^{\mathrm{DimN}}(M^{2}),
\label{eq:A1}
\end{equation}%
respectively. The components of $\rho ^{\mathrm{OPE}}(s)$ and $\Pi (M^{2})$
are given by the formulas%
\begin{equation}
\rho ^{\mathrm{DimN}}(s)=\int_{0}^{1}d\alpha \int_{0}^{1-a}d\beta \rho ^{%
\mathrm{DimN}}(s,\alpha ,\beta ),\ \ \rho ^{\mathrm{DimN}}(s)=\int_{0}^{1}d%
\alpha \rho ^{\mathrm{DimN}}(s,\alpha ),  \label{eq:A2}
\end{equation}%
and
\begin{equation}
\Pi ^{\mathrm{DimN}}(M^{2})=\int_{0}^{1}d\alpha \int_{0}^{1-a}d\beta \Pi ^{%
\mathrm{DimN}}(M^{2},\alpha ,\beta ),\ \ \Pi ^{\mathrm{DimN}%
}(M^{2})=\int_{0}^{1}d\alpha \Pi ^{\mathrm{DimN}}(M^{2},\alpha ).
\label{eq:A4}
\end{equation}%
In Eqs.\ (\ref{eq:A2}) and (\ref{eq:A4}) variables $\alpha $ and $\beta $
are Feynman parameters.

The perturbative and nonperturbative components of the spectral density $%
\rho ^{\mathrm{pert.}}(s,\alpha ,\beta )$ and $\rho ^{\mathrm{Dim3(4,5,6,7,8)%
}}(s,\alpha ,\beta )$ \ have the forms:
\begin{eqnarray}
&&\rho ^{\mathrm{pert.}}(s,\alpha ,\beta )=\frac{\Theta (L)L^{2}}{512\pi
^{6}(\beta -1)^{4}N_{1}^{4}N_{2}}\left\{ 2(\beta -1)^{2}N_{3}\left[ -\alpha
\beta N_{3}+3m_{c}m_{s}(\alpha +\beta )N_{1}^{2}\right] \right.  \notag \\
&&\left. +4(\beta -1)N_{1}^{2}L\left[ -\alpha \beta N_{3}+m_{c}m_{s}(\alpha
+\beta )N_{1}^{2}\right] -\alpha \beta N_{1}^{4}L^{2}\right\} ,
\end{eqnarray}%
\begin{eqnarray}
&&\rho ^{\mathrm{Dim3}}(s,\alpha ,\beta )=-\frac{\langle \overline{s}%
s\rangle \Theta (L)}{16\pi ^{4}(\beta -1)^{2}N_{1}^{6}}\left\{ m_{s}(\beta
-1)^{2}\alpha \beta N_{2}N_{3}^{2}+(\beta -1)N_{1}^{2}\left[ 6m_{s}\alpha
\beta N_{2}N_{3}+2m_{c}^{2}m_{s}N_{1}^{3}\right. \right.  \notag \\
&&\left. -m_{c}s\alpha \beta L(\beta ^{4}+\alpha ^{2}(\alpha -1)^{2}+\beta
^{3}(3\alpha -2)+\alpha \beta (2-5\alpha +3\alpha ^{2})+\beta ^{2}(1-5\alpha
+4\alpha ^{2}))\right]  \notag \\
&&\left. -N_{1}^{4}\left[ -3m_{s}\alpha \beta N_{2}+m_{c}L^{2}(\beta
^{3}+2\beta \alpha (\alpha -1)+\alpha ^{2}(\alpha -1)+\beta ^{2}(2\alpha -1)%
\right] \right\} ,
\end{eqnarray}%
\begin{eqnarray}
&&\rho ^{\mathrm{Dim4}}(s,\alpha ,\beta )=\frac{\langle \alpha _{s}G^{2}/\pi
\rangle \Theta (L)}{4608\pi ^{4}(\beta -1)^{2}N_{1}^{5}N_{2}^{2}}\left\{
\alpha \beta (\beta -1)^{2}\left[ -6m_{c}m_{s}(\beta ^{3}+\beta ^{2}(\alpha
-1)+\beta ^{2}\alpha +\alpha ^{2}(\alpha -1))\right. \right.  \notag \\
&&\times (3N_{2}N_{3}+2m_{c}^{2}N_{1})+N_{2}N_{3}(\alpha +\beta
)(54N_{2}N_{3}+m_{c}^{2}(\beta ^{4}-\beta ^{3}+\beta ^{2}\alpha ^{2}+\alpha
^{3}(\alpha -1))/N_{1}  \notag \\
&&+sm_{c}N_{2}^{2}\left( 11m_{c}\alpha \beta (\alpha ^{3}+\beta
^{3})+6m_{s}\left( 2\beta ^{5}-73\beta ^{3}\alpha (\alpha -1)+2\alpha
^{4}(\alpha -1)-\beta ^{4}(2+37\alpha )+\beta ^{2}\alpha \right. \right.
\notag \\
&&\left. \left. \left. \times (-36+108\alpha -73\alpha ^{2})+\beta \alpha
^{2}(-36+73\alpha -37\alpha ^{2})\right) \right) \right] +12N_{1}^{2}(\beta
-1)L\left[ 27N_{2}^{2}N_{3}\alpha \beta (\alpha +\beta )/N_{1}\right.  \notag
\\
&&+m_{c}^{2}\alpha \beta (\alpha ^{3}+\beta ^{3})N_{2}+m_{c}m_{s}\left(
2\beta ^{6}+2\alpha ^{4}(\alpha -1)^{2}-\beta ^{5}(4+35\alpha )+\beta
^{3}\alpha (-109+254\alpha -146\alpha ^{2})\right.  \notag \\
&&\left. \left. +\beta ^{4}(2+108\alpha -110\alpha ^{2})+\beta \alpha
^{2}(36-109\alpha +108\alpha ^{2}-35\alpha ^{3})-2\beta ^{2}\alpha
(-18+90\alpha -127\alpha ^{2}+55\alpha ^{3})\right) \right]  \notag \\
&&\left. +162L^{2}\alpha \beta N_{1}^{3}N_{2}(\beta ^{2}+\alpha (\alpha
-1)+\beta (2\alpha -1))\right\} ,
\end{eqnarray}%
\begin{eqnarray}
&&\rho ^{\mathrm{Dim5}}(s,\alpha ,\beta )=\frac{\langle \overline{s}%
g_{s}\sigma Gs\rangle \alpha \beta N_{2}\Theta (L)}{64\pi ^{4}(\beta
-1)N_{1}^{6}}\left\{ (\beta -1)\left[ 13m_{s}\alpha \beta
N_{2}N_{3}+4m_{c}^{2}m_{s}N_{1}^{3}-6m_{c}s\alpha \beta \left( \beta
^{4}\right. \right. \right.  \notag \\
&&\left. \left. +\alpha ^{2}(\alpha -1)^{2}+\beta ^{3}(3\alpha -2)+\beta
\alpha (2-5\alpha +3\alpha ^{2})+\beta ^{2}(1-5\alpha +4\alpha ^{2})\right)
\right] -2N_{1}^{2}L\left[ -8m_{s}\alpha \beta N_{2}\right.  \notag \\
&&\left. \left. +3m_{c}(\beta ^{3}+2\beta \alpha (\alpha -1)+\alpha
^{2}(\alpha -1)+\beta ^{2}(2\alpha -1))\right] \right\} ,
\end{eqnarray}%
\begin{eqnarray}
&&\rho ^{\mathrm{Dim6}}(s,\alpha ,\beta )=\frac{\Theta (L)}{829440\pi
^{6}(\beta -1)^{2}N_{1}^{5}N_{2}^{2}}\left\{ 3840g_{s}^{2}\langle \overline{s%
}s\rangle ^{2}\pi ^{2}(\beta -1)\alpha \beta N_{2}^{4}[s(\beta -1)\alpha
\beta N_{2}+N_{1}^{2}L]\right.  \notag \\
&&-27\langle g_{s}^{3}G^{3}\rangle m_{c}^{2}\beta ^{5}(\beta -1)^{2}\left[
2(\beta -1)\beta \alpha N_{2}-2\alpha \left( -5\beta ^{3}+\beta
^{2}(10-3\alpha )+2\alpha (\alpha -1)^{2}\right. \right.  \notag \\
&&\left. \left. \left. +\beta (-5+\alpha +4\alpha ^{2})\right) \right]
\right\} ,
\end{eqnarray}%
\begin{eqnarray}
&&\rho _{1}^{\mathrm{Dim7}}(s,\alpha ,\beta )=\frac{\langle \alpha
_{s}G^{2}/\pi \rangle \langle \overline{s}s\rangle \Theta (L)}{1152\pi
^{2}(\beta -1)N_{1}^{4}}\left\{ 90m_{s}\left( \beta -1\right) \alpha \beta
(\alpha +\beta )N_{2}^{2}+18m_{s}(\beta -1)\alpha \beta N_{2}\right.  \notag
\\
&&\times \left[ \beta ^{2}+\alpha (\alpha -1)+\beta (2\alpha -1)\right]
+8m_{c}\left[ 2\beta ^{6}+2\alpha ^{4}(\alpha -1)^{2}-\beta ^{5}(4+19\alpha
)+\beta ^{4}(2+56\alpha -37\alpha ^{2})\right.  \notag \\
&&\left. \left. +\beta ^{3}\alpha (-55+91\alpha -37\alpha ^{2})+\beta
^{2}\alpha (18-72\alpha +74\alpha ^{2}-17\alpha ^{3})+\beta \alpha
^{2}(18-37\alpha +15\alpha ^{2}+4\alpha ^{3})\right] \right\} ,
\end{eqnarray}%
\begin{equation}
\rho ^{\mathrm{Dim8}}(s,\alpha ,\beta )=-\frac{\langle \alpha _{s}G^{2}/\pi
\rangle ^{2}\alpha ^{2}\beta ^{2}N_{2}\Theta (L)}{512\pi ^{2}N_{1}^{4}}.
\end{equation}%
The function $\rho _{2}^{\mathrm{Dim7}}(s,\alpha )$ is determined by the
expression%
\begin{equation}
\rho _{2}^{\mathrm{Dim7}}(s,\alpha )=-\frac{\langle \alpha _{s}G^{2}/\pi
\rangle \langle \overline{s}s\rangle m_{c}}{288\pi ^{2}}\Theta (\widetilde{L}%
).
\end{equation}

Components of $\Pi (M^{2})$ are%
\begin{eqnarray}
&&\Pi ^{\mathrm{Dim6}}(M^{2},\alpha ,\beta )=\frac{\langle
g_{s}^{3}G^{3}\rangle m_{c}^{3}\beta ^{4}(\beta -1)}{92160M^{4}\pi
^{6}N_{1}^{6}N_{2}}\exp \left[ -\frac{m_{c}^{2}(\alpha +\beta )N_{1}}{%
M^{2}\alpha \beta N_{2}}\right] \left\{ -\left[ 2m_{c}^{5}\alpha \beta
(\beta -1)^{2}(\alpha +\beta )^{2}\right. \right.  \notag \\
&&\left. +3m_{c}M^{4}\alpha \beta N_{1}^{2}\right] \left[ 3\beta ^{3}+\beta
\alpha (\alpha -2)-\alpha ^{2}(\alpha -1)+\beta ^{2}(2\alpha -3)\right]
+6m_{s}M^{4}N_{1}^{3}\left[ 8\beta ^{3}+\beta \alpha (\alpha -3)\right.
\notag \\
&&\left. +\alpha ^{2}(\alpha -1)+\beta ^{2}(3\alpha -8)\right]
+6m_{c}^{2}m_{s}M^{2}(\beta -1)N_{1}^{2}\left[ 8\beta ^{4}+2\beta \alpha
^{2}(\alpha -2)+\alpha ^{3}(\alpha -1)\right.  \notag \\
&&\left. +\beta ^{2}\alpha (4\alpha -11)+\beta ^{3}(11\alpha -8)\right]
+6m_{c}^{4}m_{s}(\beta -1)^{2}\left[ 2\beta ^{7}-\alpha ^{5}(\alpha
-1)^{2}+\beta ^{6}(5\alpha -4)\right.  \notag \\
&&+\beta ^{3}\alpha ^{2}(2+11\alpha -14\alpha ^{2})+\beta ^{4}\alpha
(5-4\alpha -7\alpha ^{2})+\beta \alpha ^{4}(-4+9\alpha -5\alpha ^{2})+2\beta
^{5}(1-5\alpha +\alpha ^{2})  \notag \\
&&\left. -4\beta ^{2}\alpha ^{3}(1-4\alpha +3\alpha ^{2})\right]
+3m_{c}^{3}M^{2}\alpha \beta (\beta -1)\left[ 3\beta ^{6}-\beta \alpha
^{4}(\alpha -1)-\alpha ^{4}(\alpha -1)^{2}+\beta ^{5}(8\alpha -6)\right.
\notag \\
&&\left. \left. +2\beta ^{2}\alpha ^{2}(3-4\alpha +\alpha ^{2})+\beta
^{3}\alpha (8-17\alpha +8\alpha ^{2})+\beta ^{4}(3-16\alpha +11\alpha ^{2})
\right] \right\} ,
\end{eqnarray}%
\begin{eqnarray}
&&\Pi ^{\mathrm{Dim7}}(M^{2},\alpha ,\beta )=\frac{\langle \alpha
_{s}G^{2}/\pi \rangle \langle \overline{s}s\rangle m_{c}^{2}}{576\pi
^{2}M^{4}N_{1}^{7}}\exp \left[ -\frac{m_{c}^{2}(\alpha +\beta )N_{1}}{%
M^{2}\alpha \beta N_{2}}\right] \left\{ -2m_{c}M^{2}\alpha \beta (\beta -1)%
\left[ m_{c}^{2}(\alpha +\beta )(\beta -1)\right. \right.  \notag \\
&&\left. +M^{2}N_{1}\right] \left[ 2\beta ^{6}+2\alpha ^{4}(\alpha
-1)^{2}+\beta ^{5}(5\alpha -4)+\beta \alpha ^{3}(4-9\alpha +5\alpha
^{2})+\alpha ^{2}\beta ^{2}(4-13\alpha +9\alpha ^{2})\right.  \notag \\
&&\left. +\beta ^{4}(2-9\alpha +9\alpha ^{2})+\beta ^{3}\alpha (4-13\alpha
+10\alpha ^{2})\right] +m_{s}\left[ 2m_{c}^{4}(\beta -1)^{3}\alpha \beta
(\alpha +\beta )^{3}(\beta ^{3}-\beta ^{2}+\beta \alpha +\alpha ^{2}(\alpha
-1))\right.  \notag \\
&&+m_{c}^{2}M^{2}(\beta -1)^{2}\alpha \beta (\alpha +\beta )^{2}\left(
7\beta ^{5}+3\beta ^{4}(5\alpha -6)+\alpha ^{2}(\alpha -1)^{2}(7\alpha
-4)+\beta ^{3}(15-35\alpha +23\alpha ^{2})\right.  \notag \\
&&\left. +\alpha \beta (-8+28\alpha -35\alpha ^{2}+15\alpha ^{3})+\beta
^{2}(-4+28\alpha -47\alpha ^{2}+23\alpha ^{3})\right) -M^{4}N_{1}^{2}\left(
8\beta ^{7}+8\alpha ^{4}(\alpha -1)^{3}+3\beta ^{6}(3\alpha -8)\right.
\notag \\
&&+\beta ^{5}(24-30\alpha +17\alpha ^{2})+3\beta \alpha ^{3}(-4+15\alpha
-19\alpha ^{2}+8\alpha ^{3})+3\beta ^{3}\alpha (-4+12\alpha -20\alpha
^{2}+11\alpha ^{3})  \notag \\
&&\left. \left. \left. +\beta ^{4}(-8+33\alpha -45\alpha ^{2}+24\alpha
^{3})+\beta ^{2}\alpha ^{2}(-8+48\alpha -70\alpha ^{2}+33\alpha ^{3})\right)
\right] \right\} ,
\end{eqnarray}

\begin{eqnarray}
&&\Pi ^{\mathrm{Dim8}}(M^{2},\alpha ,\beta )=-\frac{\langle \alpha
_{s}G^{2}/\pi \rangle ^{2}\alpha \beta (\beta -1)}{82944\pi
^{2}M^{6}N_{1}^{8}N_{2}}\exp \left[ -\frac{m_{c}^{2}(\alpha +\beta )N_{1}}{%
M^{2}\alpha \beta N_{2}}\right] \left\{ 2m_{c}^{7}(\beta -1)^{3}\alpha
^{3}\beta ^{3}(\alpha +\beta )^{2}\right.  \notag \\
&&-6m_{c}^{6}m_{s}(\beta -1)^{3}\alpha ^{2}\beta ^{2}(\alpha +\beta
)^{3}N_{2}+81m_{c}M^{6}\alpha \beta N_{1}^{3}\left[ \beta ^{3}+\beta
^{2}(\alpha -1)+\beta \alpha ^{2}+\alpha ^{2}(\alpha -1)\right]
+216m_{s}M^{6}N_{1}^{3}  \notag \\
&&\times \left[ \beta ^{5}+\beta ^{4}(\alpha -2)-\beta ^{3}(\alpha -1)+\beta
\alpha ^{3}(\alpha -1)+\alpha ^{3}(\alpha -1)^{2}\right] +81m_{c}^{3}M^{4}(%
\beta -1)\alpha \beta N_{1}^{2}\left[ \beta ^{4}+\alpha ^{3}(\alpha
-1)\right.  \notag \\
&&\left. +\beta ^{3}(2\alpha -1)+\beta ^{2}\alpha (2\alpha -1)+\beta \alpha
^{2}(2\alpha -1)\right] +216m_{c}^{2}m_{s}M^{4}(\beta -1)N_{1}^{2}\left[
\beta ^{6}+2\beta ^{5}(\alpha -1)-\beta ^{3}\alpha (\alpha -1)\right.  \notag
\\
&&\left. +\beta ^{2}\alpha ^{3}(\alpha -1)+\alpha ^{4}(\alpha -1)^{2}+\beta
^{4}(1-3\alpha +\alpha ^{2})+\beta \alpha ^{3}(1-3\alpha +2\alpha ^{2})
\right] +54m_{c}^{5}M^{2}(\beta -1)^{2}\alpha \beta (\alpha +\beta )^{2}
\notag \\
&&\times \left[ \beta ^{5}+2\beta ^{4}(\alpha -1)+\alpha ^{3}(\alpha
-1)+\beta \alpha ^{2}(1-3\alpha +2\alpha ^{2})+\beta ^{2}\alpha (1-4\alpha
+3\alpha ^{2})+\beta ^{3}(1-3\alpha +3\alpha ^{2}\right]  \notag \\
&&-12m_{c}^{4}m_{s}M^{2}(\beta -1)^2\alpha \beta \left[ 9\beta ^{7}+9\alpha
^{4}(\alpha -1)^{3}+4\beta \alpha ^{3}(\alpha -1)(13\alpha -9)+\beta
^{6}(52\alpha -27)+\beta ^{5}\left( 27-140\alpha \right. \right.  \notag \\
&&\left. +138\alpha ^{2}\right) +3\beta ^{2}\alpha ^{2}(-18+79\alpha
-107\alpha ^{2}+46\alpha ^{3})+\beta ^{3}\alpha (-36+237\alpha -416\alpha
^{2}+215\alpha ^{3})  \notag \\
&&\left. \left. +\beta ^{4}(-9+124\alpha -321\alpha ^{2}+215\alpha ^{3})
\right] \right\} .
\end{eqnarray}

The terms $\mathrm{Dim9}$ and $\mathrm{Dim10}$ are exclusively of the type (%
\ref{eq:A4}) and have two components $\Pi _{1}^{\mathrm{DimN}}(M^{2},\alpha
,\beta )$ and $\Pi _{2}^{\mathrm{DimN}}(M^{2},\alpha )$ presented below:
\begin{eqnarray}
&&\Pi _{1}^{\mathrm{Dim9}}(M^{2},\alpha ,\beta )=\frac{m_{c}(\beta -1)}{%
17280\pi ^{4}M^{8}N_{1}^{10}}\exp \left[ -\frac{m_{c}^{2}(\alpha +\beta
)N_{1}}{M^{2}\alpha \beta N_{2}}\right] \left\{ 5\langle \alpha
_{s}G^{2}/\pi \rangle \langle \overline{s}g_{s}\sigma Gs\rangle M^{2}\pi
^{2}N_{1}^{2}N_{2}\right.  \notag \\
&&\times \left[ 2m_{c}^{5}m_{s}(\beta -1)^{3}\alpha \beta (\alpha +\beta
)^{3}\left( \beta ^{3}-\beta ^{2}+\beta \alpha +\alpha ^{2}(\alpha
-1)\right) +6M^{6}N_{1}^{3}\left( \beta ^{5}-\beta ^{4}+\alpha ^{4}(\alpha
-1)\right) \right.  \notag \\
&&+6m_{c}^{2}M^{4}(\beta -1)N_{1}^{2}\left( \beta ^{6}+\beta ^{5}(\alpha
-1)-\beta ^{4}\alpha +\beta \alpha ^{4}(\alpha -1)+\alpha ^{5}(\alpha
-1)\right) -3m_{c}^{4}M^{2}(\beta -1)^{2}\alpha \beta  \notag \\
&&\times \left( 2\beta ^{7}+2\alpha ^{5}(\alpha -1)^{2}+\beta ^{6}(7\alpha
-4)+\beta \alpha ^{4}(6-13\alpha +7\alpha ^{2})+2\beta ^{2}\alpha
^{3}(4-11\alpha +7\alpha ^{2})\right.  \notag \\
&&\left. \left. +\beta ^{5}(2-13\alpha +14\alpha ^{2})+\beta ^{3}\alpha
^{2}(8-26\alpha +19\alpha ^{2})+\beta ^{4}\alpha (6-22\alpha +19\alpha
^{2})\right) \right]  \notag \\
&&+3\langle g_{s}^{3}G^{3}\rangle \langle \overline{s}s\rangle
m_{c}^{3}(\beta -1)^{2}\left[ m_{c}M^{4}N_{1}^{4}\left( 6\beta ^{6}+8\beta
^{5}\alpha +3\beta ^{4}\alpha ^{2}+3\beta ^{2}\alpha ^{4}+8\beta \alpha
^{5}+6\alpha ^{6}\right) \right.  \notag \\
&&-8m_{s}M^{4}N_{1}^{4}\left( 2\beta ^{6}+3\beta ^{4}\alpha (\alpha
-1)+3\beta ^{2}\alpha ^{4}+3\beta \alpha ^{4}(\alpha -1)+2\alpha ^{5}(\alpha
-1)+\beta ^{5}(3\alpha -2)\right)  \notag \\
&&-m_{c}^{4}m_{s}(\beta -1)^{2}\alpha \beta (\alpha +\beta )^{3}\left(
3\beta ^{7}+2\beta ^{6}(\alpha -3)-\beta ^{3}\alpha ^{3}(\alpha -3)+\beta
^{2}\alpha ^{4}(\alpha -1)+3\alpha ^{5}(\alpha -1)^{2}\right.  \notag \\
&&\left. +\beta ^{5}(3-\alpha +\alpha ^{2})-\beta ^{4}\alpha (1+\alpha
+\alpha ^{2})+\beta \alpha ^{4}(-1-\alpha +2\alpha ^{2})\right)
-m_{c}^{2}m_{s}M^{2}(\beta -1)N_{1}^{2}  \notag \\
&&\times \left( 4\beta ^{9}-3\beta \alpha ^{7}(\alpha -1)+4\alpha
^{7}(\alpha -1)^{2}-\beta ^{8}(8+3\alpha )-3\beta ^{6}\alpha ^{2}(7\alpha
-8)+\beta ^{5}\alpha ^{2}(-12+29\alpha -25\alpha ^{2})\right.  \notag \\
&&\left. +\beta ^{4}\alpha ^{3}(-8+32\alpha -25\alpha ^{2})+\beta ^{3}\alpha
^{4}(-8+29\alpha -21\alpha ^{2})+\beta ^{7}(4+3\alpha -15\alpha ^{2})-3\beta
^{2}\alpha ^{5}(4-8\alpha +5\alpha ^{2})\right)  \notag \\
&&+m_{c}^{3}M^{2}(\beta -1)N_{1}^{2}\left( 2\beta ^{9}+\beta ^{5}\alpha
^{3}(3-5\alpha )+\beta ^{4}\alpha ^{4}(2-5\alpha )+\beta ^{3}\alpha
^{5}(3-7\alpha )+3\beta \alpha ^{7}(\alpha -1)\right.  \notag \\
&&\left. \left. \left. +2\alpha ^{8}(\alpha -1)-3\beta ^{7}\alpha (1+\alpha
)+\beta ^{8}(3\alpha -2)+\beta ^{6}\alpha ^{2}(1-7\alpha )+\beta ^{2}\alpha
^{6}(1-3\alpha )\right) \right] \right\} ,
\end{eqnarray}%
\begin{eqnarray}
&&\Pi _{2}^{\mathrm{Dim9}}(M^{2},\alpha )=\frac{\langle \alpha _{s}G^{2}/\pi
\rangle \langle \overline{s}g_{s}\sigma Gs\rangle m_{c}}{1152M^{4}\pi
^{2}\alpha ^{4}(\alpha -1)^{2}}\exp \left[ -\frac{m_{c}^{2}}{M^{2}\alpha
(1-\alpha )}\right] \left[ 2m_{c}^{3}m_{s}(\alpha -1)\right.  \notag \\
&&\left. -9m_{c}^{2}M^{2}\alpha ^{2}(\alpha -1)+9M^{4}\alpha ^{3}(\alpha
-1)^{2}+4m_{c}m_{s}M^{2}\alpha (1-\alpha -2\alpha ^{2}+3\alpha ^{3})\right] ,
\end{eqnarray}%
\begin{eqnarray}
&&\Pi _{1}^{\mathrm{Dim10}}(M^{2},\alpha ,\beta )=-\frac{m_{c}^{3}\alpha
\beta (\beta -1)^{3}}{3645\cdot 10^{11}M^{10}\pi ^{4}N_{1}^{11}N_{2}}\exp %
\left[ -\frac{m_{c}^{2}(\alpha +\beta )N_{1}}{M^{2}\alpha \beta N_{2}}\right]
\left\{ 320g_{s}^{2}\langle \overline{s}s\rangle ^{2}\pi
^{2}m_{c}^{3}M^{4}\right.  \notag \\
&&\times (\beta -1)(\beta ^{2}-\beta \alpha +\alpha ^{2})\left[ \beta
^{4}+\alpha ^{2}(\alpha -1)^{2}+\beta ^{3}(3\alpha -2)+\beta \alpha
(2-5\alpha +3\alpha ^{2})+\beta ^{2}(1-5\alpha +4\alpha ^{2})\right] ^{3}
\notag \\
&&+9\langle g_{s}^{3}G^{3}\rangle \alpha ^{2}\beta ^{2}\left[
12m_{c}^{3}M^{4}\alpha ^{2}\beta ^{2}(\beta -1)N_{1}^{3}+18m_{s}M^{6}(\alpha
+\beta )N_{1}^{5}-m_{c}^{7}\alpha \beta (\beta -1)^{3}(\alpha +\beta
)^{2}\right.  \notag \\
&&\times \left( 3\beta ^{4}+\beta ^{3}(\alpha -3)+\beta \alpha ^{2}(\alpha
-1)+3\alpha ^{3}(\alpha -1)+\beta ^{2}\alpha (2\alpha -1)\right)
-6m_{c}^{2}m_{s}M^{4}(\beta -1)N_{1}^{3}\left( 5\beta ^{4}\right.  \notag \\
&&\left. +22\beta \alpha ^{2}(\alpha -1)+5\alpha ^{3}(\alpha -1)+2\beta
^{2}\alpha (19\alpha -11)+\beta ^{3}(22\alpha -5)\right)
+3m_{c}^{6}m_{s}(\beta -1)^{3}(\alpha +\beta )^{2}\left( 5\beta ^{6}\right.
\notag \\
&&+3\beta \alpha ^{4}(\alpha -1)+5\alpha ^{4}(\alpha -1)^{2}+\beta
^{5}(3\alpha -10)-2\beta ^{3}\alpha ^{2}(7\alpha -8)+\beta ^{4}(5-3\alpha
-6\alpha ^{2})-2\alpha ^{2}\beta ^{2}  \notag \\
&&\left. \times (5-8\alpha +3\alpha ^{2})\right) +3m_{c}^{5}M^{2}\alpha
\beta (\beta -1)^{2}\left( 3\beta ^{7}+\beta ^{4}\alpha (5-4\alpha )+4\beta
^{2}\alpha ^{4}(\alpha -1)+5\beta \alpha ^{4}(\alpha -1)^{2}\right.  \notag
\\
&&\left. +3\alpha ^{5}(\alpha -1)^{2}+\beta ^{6}(5\alpha -6)+\beta
^{5}(3-10\alpha +4\alpha ^{2})\right) -3m_{c}^{4}m_{s}M^{2}(\beta
-1)^{2}\left( 13\beta ^{9}+13\alpha ^{6}(\alpha -1)^{3}\right.  \notag \\
&&+\beta ^{8}(20\alpha -39)+\beta \alpha ^{5}(\alpha -1)^{2}(20\alpha
-11)-\beta ^{2}\alpha ^{4}(\alpha -1)^{2}(45\alpha -73)+\beta
^{7}(39-51\alpha -45\alpha ^{2})  \notag \\
&&+\beta ^{4}\alpha ^{2}(73-510\alpha +808\alpha ^{2}-371\alpha ^{3})+\beta
^{6}(-13+42\alpha +163\alpha ^{2}-211\alpha ^{3})  \notag \\
&&\left. \left. \left. +\beta ^{3}\alpha ^{3}(142-510\alpha +579\alpha
^{2}-211\alpha ^{3})-\beta ^{5}\alpha (11+191\alpha -579\alpha
^{2}+371\alpha ^{3})\right) \right] \right\} ,
\end{eqnarray}%
and%
\begin{equation}
\Pi _{2}^{\mathrm{Dim10}}(M^{2},\alpha )=\frac{\langle \alpha _{s}G^{2}/\pi
\rangle g_{s}^{2}\langle \overline{s}s\rangle ^{2}m_{c}^{3}m_{s}}{%
23328M^{4}\pi ^{2}(\alpha -1)^{3}}\exp \left[ -\frac{m_{c}^{2}}{M^{2}\alpha
(1-\alpha )}\right] .
\end{equation}

In expressions above, $\Theta (x)$ is Unit Step function. We have also  used 
the following short-hand notations%
\begin{eqnarray}
N_{1} &=&\beta ^{2}+(\alpha +\beta )(\alpha -1),\ \ \ \ \ N_{2}=\alpha
+\beta -1,\ \ N_{3}=s\alpha \beta N_{2},\   \notag \\
\ \ \ L &\equiv &L(s,\alpha ,\beta )=\frac{(\beta -1)\left[
N_{3}-m_{c}^{2}(\alpha +\beta )N_{1}\right] }{N_{1}^{2}},\ \widetilde{L}%
\equiv \widetilde{L}(s,\alpha )=s\alpha (1-\alpha )-m_{c}^{2}.
\end{eqnarray}

\end{widetext}

\end{document}